\documentclass[11pt]{article}
\usepackage{amsmath,amsfonts}
\usepackage{graphicx,psfrag,epsf}
\usepackage{enumerate}
\usepackage{natbib}
\usepackage{url} % not crucial - just used below for the URL 
\usepackage{booktabs}
\usepackage{tabularx}
\usepackage{geometry}
\usepackage{lscape} 

\usepackage{zwcommands}
\usepackage{setspace}
\usepackage{mathtools}

\usepackage[algo2e,ruled,linesnumbered,vlined]{algorithm2e}
\usepackage{bm}
\usepackage{subfigure}
\usepackage{authblk}
\usepackage[format=hang,labelfont=bf]{caption}
\usepackage{url}
\usepackage{hyperref}
\usepackage{todonotes}
\usepackage[capposition=bottom]{floatrow}
\usepackage{paralist} 
\usepackage{xcolor}
\usepackage[toc,page]{appendix}

\usepackage{scalefnt}
\usepackage{bm}
\usepackage{comment}

\newcommand{\mbf}{\mathbf}
\newcommand{\squared}[1]{\left[#1\right]^2}
\newcommand{\bracsup}[1]{^{(#1)}}

\newcommand{\pau}{u}

\newcommand{\pr}{\textrm{pr}}
\newcommand{\vleaf}{\cV_{\sf \scriptsize leaf}}
\newcommand{\vstarleaf}{\cV^*_{\sf \scriptsize leaf}}
\newcommand{\pleaf}{p_{\sf \scriptsize leaf}}
\newcommand{\pstarleaf}{p^*_{\sf \scriptsize leaf}}

\makeatletter
\newcommand\makebig[2]{%
  \@xp\newcommand\@xp*\csname#1\endcsname{\bBigg@{#2}}%
  \@xp\newcommand\@xp*\csname#1l\endcsname{\@xp\mathopen\csname#1\endcsname}%
  \@xp\newcommand\@xp*\csname#1r\endcsname{\@xp\mathclose\csname#1\endcsname}%
}
\makeatother
\makebig{biggg} {3.0}
\makebig{Biggg} {3.5}
\makebig{bigggg}{4.0}
\makebig{Bigggg}{4.5}

\providecommand{\keywords}[1]
{
  \small	
  {\textit{Keywords:}} #1
}

\begin{document}
	\title{\bf Tree-informed Bayesian multi-source domain adaptation: cross-population probabilistic cause-of-death assignment using verbal autopsy}
		\author[1,2]{Zhenke Wu}
		\author[3]{Zehang Richard Li}
		\author[1]{Irena Chen}
		\author[1]{Mengbing Li}
		\affil[1]{Department of Biostatistics, University of Michigan, Ann Arbor, MI 48109, USA}
		\affil[1]{Michigan Institute for Data Science, Ann Arbor, MI 48109, USA}
		\affil[3]{Department of Statistics, University of California, Santa Cruz, CA 95064, USA}
		\date{}
		\maketitle
\vspace{-1cm}
\begin{center}
%{\sf \textbf{Version}: 0.97\\
% \textbf{Compiled}: \date{\today}}
\end{center}

\abstract{

Determining causes of deaths (COD) occurred outside of civil registration and vital statistics systems is challenging. A technique called verbal autopsy (VA) is widely adopted to gather information on deaths in practice. A VA consists of interviewing relatives of a deceased person about symptoms of the deceased in the period leading to the death, often resulting in multivariate binary responses. While statistical methods have been devised for estimating the cause-specific mortality fractions (CSMFs) for a study population, continued expansion of VA to new populations (or ``domains") necessitates approaches that recognize between-domain differences while capitalizing on potential similarities. In this paper, we propose such a domain-adaptive method that integrates external between-domain similarity information encoded by a pre-specified rooted weighted tree. Given a cause, we use latent class models to characterize the conditional distributions of the responses that may vary by domain. We specify a logistic stick-breaking Gaussian diffusion process prior along the tree for class mixing weights with node-specific spike-and-slab priors to pool information between the domains in a data-driven way. Posterior inference is conducted via a scalable variational Bayes algorithm. Simulation studies show that the domain adaptation enabled by the proposed method improves CSMF estimation and individual COD assignment. We also illustrate and evaluate the method using a validation data set. The paper concludes with a discussion on limitations and future directions. 

% An R package \verb"doubletree" is free and publicly available at \url{https://github.com/zhenkewu/doubletree}.
}


\section{Introduction}
\label{sec:intro}

\subsection{Verbal Autopsy (VA): background}

Patterns of mortality and causes of death at the community level are critical to informing public health policies, tracking trends, and prioritizing interventions for local governments and public health officials. Civil registration systems that track births, deaths and their causes provide the basis for countries to identify their most pressing health issues.
Despite ongoing global efforts to strengthen the civil registration and vital statistics (CRVS) system,  two-thirds of 56 million annual deaths go unrecorded, especially in low- and middle-income countries (LMIC), leaving glaring gaps for reliable mortality information \citep{world2021civil}.  
% To address the issue, often dubbed as ``invisible deaths", global efforts have been invested in 
Verbal autopsy (VA) is one of the most well-established and realistic methods to collect information about cause of death (COD) in these situations when medically certified cause of death is unavailable.
% has been established and adopted by countries lacking CRVS as a realistic and practical complementary alternative. 
VAs collect information on deaths by interviewing caregivers (or individuals who witnessed the death) of the deceased. Typically, information about healthcare access, demographic information, and various indicators of symptoms leading to the death are collected. See \citet{chandramohan2021estimating} for a recent review of historical developments, ongoing efforts to standardize VA instruments and implications for LMIC.

The central analytic goal is to use VA data to derive population-level cause-specific mortality fractions (CSMFs) and to produce individual-level COD assignment. In particular, VA data contain pertinent information on signs, symptoms, and circumstances leading to death, generically referred to as ``symptoms" in this paper. These symptoms are often coded into binary ``Yes"/``No" answers, resulting in data sets that contain multiple binary responses for each death. 
Algorithmic and probablistic methods have been developed to automate the task of estimating CSMFs and individual-level COD assignment. 
% Validation data sets with gold-standard cause-of-death labels coded by trained physicians have been curated with the goal of methods comparison and improvements \citep{murray2011population}. 
See \citet{li2021openva} for an excellent comprehensive review of the major methods and software as well as the references therein for recent methodological improvements. Developments in analytic methods and reproducible open-source software for VA have greatly fostered confidence in large-scale implementations of VA in many LMICs.

\subsubsection{Statistical challenges in domain adaptation in  VA research}

However, one emerging analytical challenge in expanding VA to new populations is in developing approaches that recognize and address potential differences between the existing and new populations in terms of the joint distribution of causes of death and VA responses. Such a problem is an example of ``domain adaptation" in machine learning literature \citep[e.g.,][]{pan2009survey}, but has received relatively little attention in VA research. In particular, CSMFs comprise a vector of population-level marginal probabilities of the causes and may differ by domain; this is most natural because a cause may differentially contribute to deaths 
occurred in different study populations. The conditional distribution of the VA responses given a cause characterizes symptom-cause relationships and may also differ by domain; we focus our paper and model formulation on addressing this aspect of domain adaptation.

It is well known that more accurate estimation of the conditional distribution of the multivariate binary VA responses given a cause can result in substantial improvements to CSMF estimation performance \citep[e.g.,][]{kunihama2020bayesian, li2020using}. To acknowledge potential between-domain differences in these conditional distributions given a cause, it is therefore tempting to directly estimate them for each domain separately. However, in a domain with few sampled deaths due to a cause, such direct estimates are often vulnerable to statistical instability, hindering accurate CSMF estimation. This issue worsens still if for that same cause the numbers of sampled deaths are small in multiple domains. In such cases,  pooling information from similar domains would improve the estimation of these conditional distributions which in turn would propagate to improving the estimation of CSMFs. On the other hand, an extreme complete-pooling approach that forces domains to have an identical conditional distribution of the VA symptoms given any cause is restrictive and would obscure the study of important between-domain variations in response patterns \citep[e.g.,][]{king2008verbal,mccormick2016probabilistic}. Data-driven pooling of information between the domains for each cause  is desirable.

\subsubsection{Existing literature}
\label{sec:literature}
Here we briefly describe a few existing work related to domain adaptation in verbal autopsy studies and how the proposed method differ from them. \citet{datta2020regularized} and \citet{fiksel2021generalized} developed methods that calibrate CSMF estimates obtained from VA algorithms trained on a training data set to produce CSMF estimates in a new population. These calibration methods differ from our work in three important ways. First, such calibration methods only consider the estimated CSMFs from a list of trained VA algorithms, and are hence not designed for using individual-level information in the training data set to perform calibration.  Second, the calibration relies on a small number of deaths with medically-confirmed causes in the new population. Third, causes often need to be manually combined before calibration can be applied to produce stable and meaningful results. Our work focus on using all individual-level data from multiple populations (referred to as ``source domains'') with known causes of death, and cause-of-death assignment in a new population (referred to as ``target domain''). The proposed method does not require known cause-of-death labels in the target domain or any ad hoc collapsing of causes. 
% In this paper, we precisely address these three differences in the assumed data setup and provide a domain-adaptive method that 1) uses individual-level VA data from multiple populations with observed causes of death (referred to as ``source domains") to conduct CSMF estimation and probabilistic cause-of-death assignment in a new population (also referred to as ``target domain"), 2) works even if the target domain may not have any gold-standard cause-of-death label at all, and 3) removes the reliance upon ad hoc collapsing of causes.

Among a few methods that directly model the individual-level VA data under domain adaptation, one related work is \citet{moran2021bayesian} that introduces a factor regression method to let the conditional distribution of the VA symptoms given a cause vary by additional individual-level covariates, which may include dummy domain indicators. This approach again is not designed for the scenario where cause-of-death labels are fully unobserved in the target domain. Our work is most related to \citet{li2021bayesian}, where a latent class model framework was proposed to model the conditional dependence relationship among symptoms with improved interpretability and computational speed. However VAs collected from different domains are treated as marginally independent data sets. Our work significantly extends \citet{li2021bayesian} by proposing a framework that can integrate additional external structural information across the domains. This allows us to efficiently pool information across domains and improve the stability of the estimates when some causes are rarely observed.

\subsection{Main contributions}

This paper develops a novel tree-integrative framework for CSMF estimation and individual-level COD assignment based on latent class models \citep{lazarsfeld1950thelogical} that jointly models multivariate binary data obtained from multiple source domains and a target domain. Our framework explicitly acknowledges  domain-by-domain variation in the distribution of causes and distribution of symptoms given causes. Most importantly, it takes into account the structural similarities among deaths from related domains. Our main methodological contributions are two folds.

% CSMFs in the target domain where causes of deaths are not observed. At the individual level, in the source domain data, we observed VA survey responses, causes, domain indicators; in the target domain data, we observe all but the causes. 

% First, in a latent class model framework \citep{Goodman1974}, we model the conditional dependence structure given each cause, in the source and target domains. However, we explicitly acknowledge potential domain-by-domain variation in the strength and direction of the covariations among the cause-specific conditional dependence. The CSMFs may also vary by domain. 

First, we propose a data-driven pooling of information across domains via a pre-specified hierarchy represented by a rooted weighted tree. This approach is shown to encourage similar conditional dependence structure across domains while recognizing important between-domain differences resulting in more accurate CSMF estimation. Simulation studies show the proposed approach has better performance compared to estimates that either completely, incorrectly, or minimally pool information across domains. Although the method is general, in this paper, we illustrate the method by a domain tree defined by geographical region of each study site, which serves as a proxy for potential regional variations in factors which may drive differences in the conditional distributions of symptoms given a cause, e.g., VA interviewer training, culture in symptom disclosure of a deceased, etc. As a secondary feature, the proposed approach also uses a hierarchy over the causes to enable information pooling between causes so that a rare cause can be pooled with similar causes to produce more stable estimates of the class-specific response profiles.

Second, we propose a tree-based measure of dissimilarity in symptom-cause relationship between the target domain and each of the source domains, separately for each cause. The proposed measure admits rich interpretation of empirical evidence about the manner in which causes differ in between-domain similarities. For example, causes with highly recognizable and specific symptoms (e.g., ``Drowning") may have symptom-cause conditional distributions that remain similar regardless of the domains, while less so for other causes with complex etiologies that are prone to differential reporting patterns across the domains.

% Computational speed. \zw{need testing other methods; our model is also a joint modeling approach instead of two-stage approaches.}

\paragraph{Paper organization} The rest of the paper is organized as follows. Section \ref{sec:model} reviews tree-related terminologies and presents the proposed model. Section \ref{sec:prior_doubletree} specifies prior distributions. A variational Bayes algorithm is presented in Section \ref{sec:algorithm}. Section \ref{sec:simulation} conducts simulation studies to illustrate the operating characteristics of the proposed method. In Section \ref{sec:dataapp}, we use a validation data set to illustrate the method. The paper concludes with a brief summary and discussion on limitations and some future directions.

\section{Model}
\label{sec:model}

We first introduce necessary terminologies and notations for characterizing a rooted weighted tree. The proposed nested latent class model (NLCM) is then formulated for deaths, each with an observed link to a leaf in a tree over source and target domains. 

\subsection{Rooted weighted trees}

A rooted tree is a graph $\cT = (\cV, E)$ with node set $\cV$ and edge set $E$ where there is a root $u_0$ and each node has at most one parent node. Let $p =  |\cV|$ represent the total number of leaf and non-leaf nodes. Let $\vleaf \subset \cV$ be the set of leaves (i.e., nodes without children), and $\pleaf = |\vleaf| < p$. We typically use $u$ to denote any node ($u \in \cV$) and $v$ to denote any leaf ($v \in \vleaf$). Each edge in a rooted tree defines a \textit{clade}: the group of leaves below it. Splitting the tree at an edge creates a partition of the leaves into two groups. For any node $u \in \cV$, the following notations apply: $c(u)$ is the set of offspring of $u$, $pa(u)$ is the parent of $u$, $d(u)$ is the set of descendants of $u$ including $u$, and $a(u)$ is the set of ancestors of $u$ including $u$. At the top of Figure \ref{fig::double_tree_scheme}, a hypothetical tree for $G=4$ source domains and one target domain with $p=8$ and $\pleaf=5$ is shown. If $u$ = 2, then $c(u) = \{5,6\}$ , $pa(u)$ = \{1\}, $d(u) = \{ 2,5,6 \}$, and $a(u) = \{1,2\}$. See Figure \ref{fig:twotrees_deathcounts} (top margin) for an instance of a nested hierarchy for six domains where VA data are collected: $\pleaf=6$ leaves representing six study sites, and $p-\pleaf=3$ non-leaf nodes subsuming the six leaf descendants (root node representing ``global"; two internal nodes representing two countries, ``India" and ``Tanzania").

Edge-weighted graphs appear as a model for numerous problems where nodes are linked with edges of different weights. In particular, the edges in $\cT$ are attached with weights where $w: E\rightarrow \RR^+$ is a weight function.  Let $\cT_w = (\cT, w)$ be a rooted weighted tree. A path in a graph is a sequence of edges which joins a sequence of distinct vertices. For a path $P$ in the tree connecting two nodes, $w(P)$ is defined as the sum of all the edge weights along the path, often referred to as the ``length" of $P$. The distance between two vertices $u$ and $u'$, denoted by $dist_{\cT_w}(u,u')$ is the length of a shortest (with minimum length) $(u,u')$-path. $dist_{\cT_w}$ is a distance: it is symmetric and satisfies the triangle inequality. In this paper, we use $w_{\pau}$ to represent the edge length between a node $u$ and its parent node $pa(u)$. $w_u$ is fully determined by $\cT_w$. For the root $u_0$, there are no parents, i.e. $pa(u_0)=\emptyset$; we set $w_{u_0}=1$. In VA contexts, although $w_u$ may be specified via the dendrogram resulting from a hierarchical clustering of domain-level covariates, in Section \ref{sec:dataapp} we will set $w_u=1$ to use minimal external domain similarity information (geographical region) for simpler exposition.

%The rooted weighted tree we will use in data analysis (Section \ref{sec:dataapp}) is a nested hierarchy of six study sites in a validation data set for a total of $N=7,841$ deaths, where the $p_L = |\vleaf|=6$ leaves represent the six sites and the $p - p_L = 3$ internal (non-leaf) nodes represent countries subsuming the observed leaf descendants, e.g., India and Tanzania each has two children leaves. 

\subsection{Nested latent class models (Nested LCM)}

Although LCMs work for multiple discrete responses of more than two levels \citep[e.g.,][]{lazarsfeld1950thelogical}, in this paper, we present the model for multivariate binary responses for simpler exposition.

\paragraph{Notations}  Let $\bX_i = (X_{i1}, \ldots, X_{iJ})^\transp\in \{0,1\}^J$ be a vector of $J$ binary responses for subject $i\in [N]$ where $N$ is the total number of subjects; here $[Q]=\{1, \ldots, Q\}$ generically represents positive integers no greater than a positive integer $Q$. Let $(Y_i,D_i)$ represent (cause of death, domain), where $Y_i$ takes its value from $\{1,\ldots, C\}$ indicating the cause of death among a total of $C$ pre-specified causes; let $\bY=(Y_1, \ldots, Y_N)^\transp$. $D_i$ takes its value from $\{0,1, \ldots, G\}$ indicating subject $i$'s domain membership: $0$ for target domain, and $1$ to $G$ for $G$ pre-specified source domains.  Let $\bD = (D_1, \ldots, D_N)^\transp$.  Throughout the paper, $D_i$ is assumed to be observed for all subjects; $Y_i$ is assumed to be observed for subjects in the source domain $\{i: D_i\neq 0\}$ but unobserved for subjects in the target domain $\{i: D_i=0\}$. Let $\bY^{\sf obs} = \{Y_i: D_i\neq 0\}$ and $\bY^{\sf mis} = \{Y_i: D_i= 0\}$; we then have $\bY=(\bY^{\sf obs},\bY^{\sf mis})^\transp$. Let $\Xb = \left(\bX_1, \ldots, \bX_N\right)^\transp$ be an $N\times J$ binary data matrix for all subjects. $\bD$ maps every row of data $\Xb$ to a leaf in the tree for domains $\cT_{w}$. Similarities between domains are then characterized by between-domain distances in $\cT_w$. Finally, let $\cD=(\Xb,\bY^{\sf obs},\bD)$ represent the data from all the domains. We use ``$\pr(A\mid B)$" to represent the conditional density of random variable(s) in $A$ given  $B$.

% Finally, let $\Xb^\ast=2\Xb-1$ be element-wise recoding of VA responses from $0$ in $\Xb$ to $-1$ in $\Xb^\ast$.

% In summary, five pieces of information are used as model inputs: 1) VA multivariate binary responses $\Xb$, 2) domain memberships $\bD$, 3) causes-of-death in the source domains $\bY$, 4) a rooted weighted tree for domains $\cT_w$, and 5) a rooted weighted tree for causes $\cT^*_{w^*}$.

\subsubsection{Model formulation} 
We assume the following model specifications for $\cD$:
\begin{align}
{\sf cause~of~death}:~ & Y_i \mid D_i=g  \sim {\sf Categorical}_C(\bpi^{(g)}), \label{eq:csmf}\\
{\sf latent~class}: ~ & Z_{i} \mid Y_i = c, D_i=g \sim {\sf Categorical}_K(\blambda^{(c,g)}),\label{eq:nlcm_class}\\
{\sf responses}:~& X_{ij} \mid Z_{i}=k,Y_{i}=c  \overset{\sf indep.}{\sim} {\sf Bernoulli}(\theta_{jk}^{(c)}), j\in [J]\label{eq:nlcm_response}
\end{align}
for $i\in[N]$, $g\in\{0\}\cup [G]$, where the population parameters $\bpi^{(g)}=(\pi_1^{(g)},\ldots, \pi_c^{(g)})^\transp$ with $\sum_{c=1}^C\pi_c^{(g)}=1$ are referred to as ``cause-specific mortality fractions" (CSMFs). Importantly, $\{\bpi^{(g)},g=0,1,\ldots, G\}$ are not constrained to be identical. We seek to estimate $\bpi^{(0)}$ and $\{Y_i: D_i=0\}$.

On the latent classes, $\bZ=(Z_1, \ldots, Z_N)^\transp\in[K]^N$ is a vector of class memberships for all subjects; $\blambda^{(c,g)} = (\lambda_1^{(c,g)},\ldots,\lambda_K^{(c,g)})^\transp$ is a vector of class weights that sum to one: $\sum_{k=1}^K\lambda_k^{(c,g)}=1$. See Remark \ref{remark:latentclass} for the nuisance role of $\bZ$ and the nuisance notion of ``class" that are introduced for inducing conditional stochastic dependence among VA responses given any pair of cause and domain. On terminology, by Equation (\ref{eq:nlcm_class}), the $K$ latent classes that $Z_i$ can take are nested within a cause of death $c$; we refer to the model as ``nested" latent class model.

On the class-specific response probabilities, for a subject died of cause $c$, $\theta_{jk}^{(c)}\in [0,1]$ is the positive response probability for item $j$ in class $k$. Let $\btheta_{\cdot k}^{(c)}=(\theta_{1k}^{(c)},\ldots, \theta_{Jk}^{(c)})^\transp$ be the vector of the $k$-th class response probability profile; let $\bTheta^{(c)}=\left(\btheta_{\cdot 1}^{(c)},\ldots,\btheta_{\cdot K}^{(c)}\right)$ collect these probabilities into a $J\times K$ matrix with $(j,k)$-th element $\theta_{jk}^{(c)}$. Note that $\bTheta^{(c)}$ may vary by cause $c$. This admits flexible characterization of symptoms that may have distinct distributions for different true causes-of-death. In addition, for any given $c$, $\bTheta^{(c)}$ is assumed domain-invariant (source or target) to facilitate shared interpretations. Conversely, letting the class response profiles vary greatly between domains would weaken the diagnostic explanability of the VA questionnaire items. See Figure \ref{fig::double_tree_scheme} for a schematic representation of the data generating process under the proposed model.

\begin{figure}[h]
\centering\captionsetup{format=plain,width=0.9\textwidth}
\includegraphics[width=.9\textwidth]{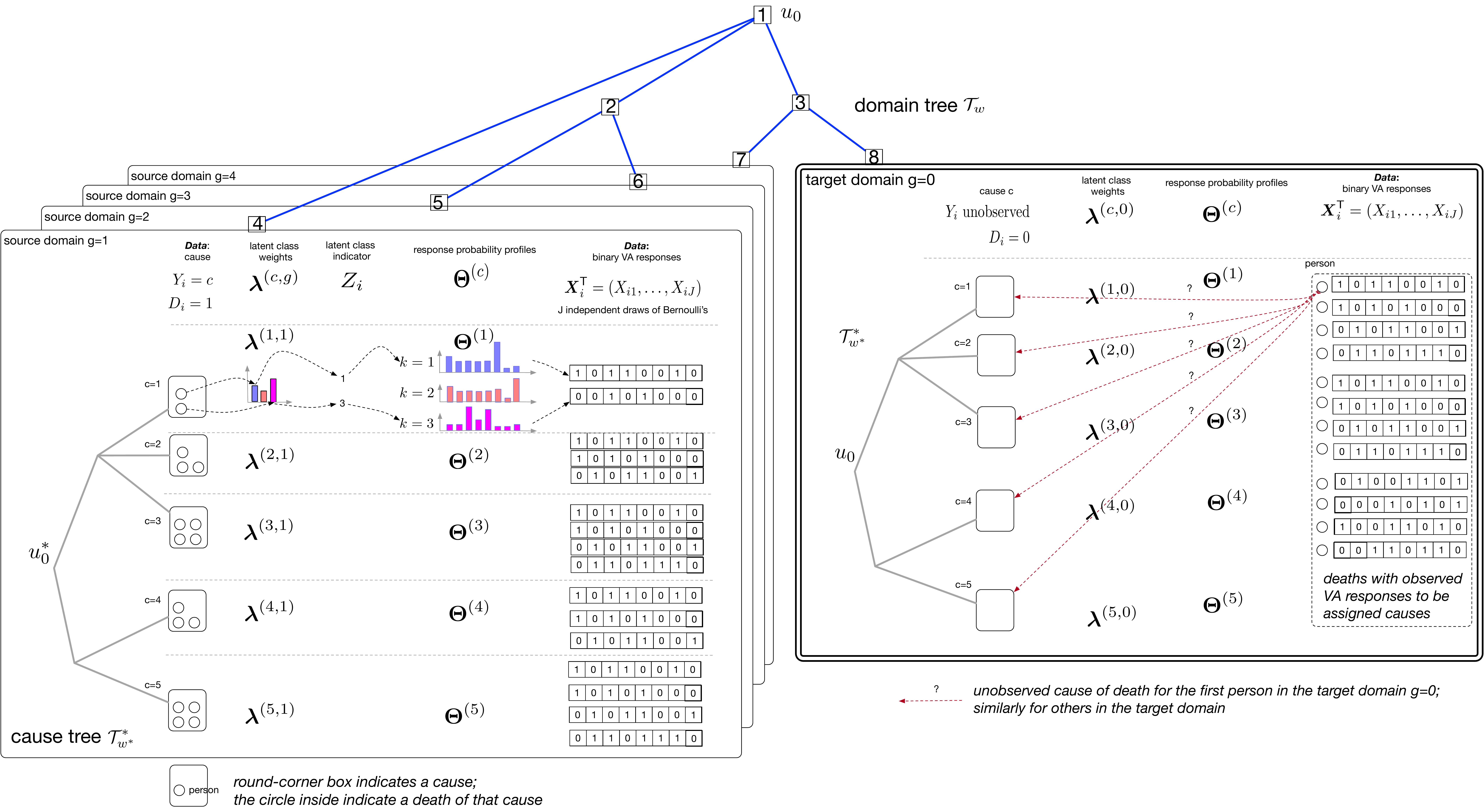}
\caption{Schematic representation of the nested LCM model structure. {\it Top}) An eight-node (root $u=u_0$) tree over five hypothetical domains is used to specify a tree-structured shrinkage prior for $\blambda^{(c,g)}, g=0, 1, \ldots, G$; {\it Left}) $G=4$ source domains ($D_i=1,...,4$), shown in overlaid plates; causes of deaths are observed to be in one of $C=5$ hypothetical causes. Hypothetical observed $J=8$ binary VA responses are also shown; {\it Right}) one target domain where the causes of deaths $Y_i$'s are unobserved but the binary VA responses $\bX_i$'s are observed. The cause hierarchy, as a secondary feature, is represented by a tree with five leaves representing $C=5$ causes. Three latent classes ($K=3$) are illustrated here.} 
\label{fig::double_tree_scheme}
\end{figure}

For any given cause $c$, despite shared $\bTheta^{(c)}$ across the domains, $\pr(\bX_i\mid Y_i=c,D_i=g)$ may differ by domain $g$ as a result of distinct class weights $\blambda^{(c,g)}$ across the domains. This is readily seen from Equations (\ref{eq:nlcm_class}) and (\ref{eq:nlcm_response}) which imply that the conditional distributions are fully parameterized by $(\bTheta^{(c)},\blambda^{(c,g)})$:
\begin{align}
    \pr(\bX_i \mid Y_i=c,D_i=g) = \sum_{k=1}^K\lambda_k^{(c,g)}\cdot\prod_{j=1}^J\left\{\theta_{jk}^{(c)} \right\}^{X_{ij}}\left\{1-\theta_{jk}^{(c)} \right\}^{1-X_{ij}},g=0,1,\ldots,G. \label{eq:conddist}
\end{align}
If $\blambda^{c,g}={\blambda^{(c,g')}}$ for any $g,g'=0,1,\ldots, G$, Equation (\ref{eq:conddist}) simplifies to $\pr(\bX_i \mid Y_i=c)$.

\begin{remark}
\label{remark:latentclass}
The model treats $\bZ$ and the notion of ``class" as technical nuisances that are introduced for the sole purpose of flexibly modeling $\pr(\bX_i \mid Y_i,D_i)$ \citep{dunson2009nonparametric}. In particular, when $K\geq 2$ and $\{\Theta^{(c)}\}$ differ by $c$, although Equation (\ref{eq:nlcm_response}) assumes conditional independence given a latent class and a cause, by integrating over $Z_i$ with probabilities $\blambda^{(Y_i,D_i)}$, we induce stochastic dependence among the $J$ components of $\bX_i$ (Equation (\ref{eq:conddist})). Setting $K=1$ would assume that the VA responses are mutually independent given any pair of cause and domain. 
\end{remark}

\begin{remark}
Equations (\ref{eq:csmf}) to (\ref{eq:nlcm_response}) under $D_i=0$ are equivalent to a $K\cdot C$-class LCM for $\{\bX_i: D_i=0\}$ parameterized by $(\bpi^{(0)},\blambda^{(c,0)},\bTheta^{(c)},c\in[C])$. Fortunately, based on data from the source domains, the multivariate binary VA response data $\{\bX_i:D_i\neq 0\}$ tabulated by the observed causes of deaths ($\bY^{\sf obs}$) provide direct information for estimating $\bTheta^{(c)}$ that is shared across the domains. 

\end{remark}

% \begin{remark}
% The class-specific response probability profiles represent typical patterns in subjects who died of a specific cause. It is preferable to maintain their interpretability by sharing them across the domains, so what a cause would manifest in terms of symptoms in survey responses would be ``indicative" or ``specific" to that cause. Alternatively, letting the class response profiles vary greatly between domains would weaken the diagnostic explanability of each survey item. $\bTheta^{(c)}$ may vary by cause. This is most natural because the constellation of symptoms may differentially appear depending on the true cause of a death.
% \end{remark}

\section{Priors}

\subsection{Tree-structured shrinkage prior}
\label{sec:prior_doubletree}

\subsubsection{Motivation and overview}
\label{sec:doubletree_specialization}

Estimating target domain CSMFs and individual cause-of-death assignment rely on efficient learning of $\pr(\bX_i \mid Y_i=c,D_i=0)$, the joint distribution of multivariate binary VA responses given each cause. In particular, by Bayes rule
\begin{align}
    \PP(Y_i=c \mid \bX_i,D_i=g) = \frac{\pr(\bX_i \mid Y_i=c,D_i=g)\pi_c^{(g)}}{\sum_{c'=1}^{C}\pr(\bX_i \mid Y_i=c',D_i=g)\pi_{c'}^{(g)}},g=0,1,\ldots,G.
\end{align}
When $g=0$, even with known $\bpi^{(0)}$, a poor estimate of $\pr(\bX_i \mid Y_i=c,D_i=0)$ can adversely impact the cause-of-death assignment on an individual level; when $\bpi^{(0)}$ is unknown as in our context, a good estimate of the conditional distribution $\pr(\bX_i \mid Y_i=c,D_i=0)$ remains critical to stable statistical estimation, e.g., via expectation-maximization for finite mixture models. The same issue persists for accurate individual-level cause-of-death assignment when $g\neq 0$ for which $\bpi^{(g)}$ can be directly estimated. However, obtaining a good estimate of $\pr(\bX_i \mid Y_i=c,D_i=g)$ is often challenging when there exist small or even zero cell counts for particular combinations of $(c,g)$, which renders direct estimation of $\pr(\bX_i \mid Y_i=c,D_i=g)$ statistically unstable if not impossible. 

This motivates us to take advantage of potential between-domain similarities. We achieve this aim by learning $G+1$ conditional distributions $\pr(\bX_i \mid Y_i=c,D_i=g),g=0,1,\ldots, G$, in a data-driven way for causes $c=1,\ldots, C$, respectively. Consider any cause $c$, by Equation (\ref{eq:conddist}), $\pr(\bX_i \mid Y_i=c,D_i=g)$ is fully parameterized by $(\bTheta^{(c)},\blambda^{(c,g)})$. Because $\bTheta^{(c)}$ is shared across domains, the $G+1$ vectors of class-mixing weights $\{\blambda^{(c,g)},g=0,1,\ldots,G\}$ fully determines the between-domain differences in Equation (\ref{eq:conddist}). This points us to the strategy of encouraging \textit{a priori} similar values of vectors $\blambda^{(c,g)}$ across domain $g=0,1,2,\ldots, G$; the degree to which they are similar may differ by cause. Data from all the domains can then be used to learn the degrees of optimal pooling between the domains for the $C$ causes, respectively. 

To achieve this aim, in Section \ref{sec:domain_hieararchy}, we introduce a tree-structured shrinkage prior for the $G+1$ vectors of class-mixing weights $\{\blambda^{(c,g)},g=0,1,\ldots,G\}$ for each $c$. The prior is based on a logistic stick-breaking Gaussian process, diffused along a pre-specified rooted weighted domain tree with $G+1$ leaves that encodes external between-domain similarity information.  Also see \ref{sec:tree_shrinkage_overview} in the Supplementary Materials for a review of the general statistical strategy of specifying tree-structured shrinkage priors \citep{thomas2019estimating,Li2021_treelcm} .

\subsubsection{\bf Logistic stick-breaking Gaussian diffusion prior for $\blambda^{(c,g)}$: integrating domain hierarchy} 
\label{sec:domain_hieararchy}

Recall that $\cT_w = (\cT=(\cV, E),w)$ represents a rooted weighted tree over domains, where the $G+1$ leaves $\vleaf$ comprise $G$ source domains and one target domain; see Section \ref{sec:databackground} for an example in the context of VA. Each domain comprises multiple independent observations $\{(\bX_i,Y_i): D_i=g\}$. Each domain $g$ is one-to-one mapped to a leaf in $\cT_w$. We specify a prior based on a logistic stick-breaking Gaussian process diffused along $\cT_w$ and end at $G+1$ leaves, inducing a prior distribution over the class weights $\blambda^{(c,g)},g=0,1,\ldots,G$.  We first reparameterize $\blambda^{(c,g)}$ with a stick-breaking representation: $\lambda_{k}^{(c,g)} = V_{k}^{(c,g)}\prod_{s < k}(1-V_{s}^{(c,g)})$, for $k\in[K]$, where $0 < V_{k}^{(c,g)} < 1$, for $k\in [K-1]$ and $V_{K}^{(c,g)} = 1$. In particular, let $\eta_{k}^{(c,g)} = \sigma^{-1}(V_{k}^{(c,g)})$, $k\in [K-1]$, $g\in \vleaf$, where $\sigma(x) = 1/\{1+\exp(-x)\}$ is the sigmoid function. The logistic stick-breaking parameterization is completed by 
\begin{equation}
\lambda_{k}^{(c,g)} = \{\sigma(\eta_{k}^{(c,g)})\}^{\mathbf{1}\{k<K\}}\prod_{s < k}\sigma(-\eta_{s}^{(c,g)}), k\in [K],\label{eq:logitSB}
\end{equation}
where $\mathbf{1}\{A\}$ is indicator function and equals 1 if statement $A$ is true and 0 otherwise. This reparametrization lends itself to simple and accurate posterior inference via variational Bayes algorithms.

For a leaf $g\in \vleaf$, let
\begin{align}
\eta_{k}^{(c,g)} & = \sum_{u\in a(g)}\xi_{k}^{(c,u)}, k\in [K-1]. \label{eq:tree_shrinkage_domain}
\end{align}
Note that $\eta_{k}^{(c,g)}$ is defined for leaves $g=\{0\}\cup[G]$ only and $\xi_{uk}^{(c,g)}$ is defined for all the nodes $u\in \cV$. Finally, for $c=1,\ldots, C$, we specify
\begin{align}
\xi_{k}^{(c,u)} & =  s_{cu} \alpha_{k}^{(c,u)}, \forall ~u\in \cV,\label{eq:ss_lambda}\\
\alpha_{k}^{(c,u)}  & \sim N(0,\tau_{\ell_{u}}w_u), ~\text{independently~for}~k\in [K-1], \forall ~u\in \cV, \label{eq:alpha_norm}\\
s_{cu_0}=1, \text{~and~} s_{cu} & \sim {\sf Bernoulli}(\rho_{c\ell_{u}}), ~\text{independently~for}~u\in \cV\setminus u_0, \label{eq:ssindicator}\\
\rho_{c\ell} & \sim {\sf Beta}(a_{c\ell}, b_{c\ell}), \textrm{~independently~for~}\ell \in [L], \label{eq::hyper_prior_with_level}
\end{align}
where $N(m',s')$ represents a Gaussian  with mean $m'$ and variance $s'$. In addition, $\ell_u\in[L]$ and maps node $u\in\cV$ (leaf or non-leaf) to one of $L$ levels; this enables distinct degrees of diffusion for $L$ non-overlapping blocks of a pre-specified partition of the nodes. Let $\btau = (\tau_{1}, \ldots, \tau_{L})^\transp$ be the diffusion variances for $L$ levels of nodes in the domain tree. Let $\bs = \{s_{cu},c\in[C],u\in \cV\}$ be a $C\times p$ matrix of slab component indicators. Let $\brho$ be a $C\times L$ matrix with $(c,\ell)$-th element $\rho_{c\ell}$. Note that the spike-and-slab indicator probabilities $\rho_{c\ell_u}$ for a node $u$ in the domain tree may vary by cause; relative to a more restrictive form $\rho_{c\ell_u}=\rho_{\ell_u}$, the present specification has the additional flexibility of cause-specific degree of pooling between domains. When $K>1$ and $s_{cu}=1, \forall u\in \vleaf$, we would assume domains differ in class weights with probability one.

\subsection{Prior for other parameters}

We assume independent Dirichlet priors for the CSMFs in the source and target domains:
\begin{align}
\bpi^{(g)} \overset{d}{\sim} {\sf Dirichlet}(\bd^{(g)}), g=0,1,\ldots, G, \label{eq:cmsf_dirichlet}
\end{align}
 where $\bd^{(g)} = (d^{(g)}_1, \ldots, d^{(g)}_C)^\transp$ is a vector of hyperparameters; let $\bd = \{\bd^{(g)}: g=0,1,\ldots, G\}$. In our simulations, we simply use $\bd^{(g)}=\mathbf{1}$ to represent a uniform prior over all cause which works well empirically. In practice, informative knowledge can be incorporated by modifying these $\bd^{(g)}$ hyperparameters to match prior numbers of observed deaths of each cause in domain $g$.

\subsubsection{Secondary feature of the method: integrating cause hierarchy} 
\label{sec:cause_hierarchy}

In a particular analysis, the number of causes can be large. Causes considered to be similar in nature and etiology would produce a symptom with similar probabilities. In addition, the number of deaths due to a cause can be small, resulting in unstable estimation of the response profiles if done separately from other causes. To overcome these issues, $\{\bTheta^{(1)},\ldots, \bTheta^{(C)}\}$ is also equipped with a tree-structured prior with a pre-specified cause tree of $C$ leaves that encodes between-cause similarities. For example, Figure \ref{fig:twotrees_deathcounts} shows a cause tree in our VA application (left margin) representing a hierarchy of $\pstarleaf=35$ causes and $p^*-\pstarleaf=7$ internal nodes that represent coarser aggregated causes; we specify all edge weights to be one. By doing so, we encourage \textit{a priori} similar values of $\bTheta^{(c)}$ across causes $c=1,\ldots, C$. This facilitates optimal pooling of information over causes and overcomes statistical stability issues for rare causes that would otherwise require ad hoc manual cause aggregations \citep{datta2020regularized}. Although the proposed approach can accommodate two hierarchies (domain and cause), because domain adaption is our primary goal, in the following we will focus empirical evaluations on the use of domain hierarchy.

Let $\cT_{w^*}^*=(\cT^*=(\cV^*,E^*),w^*)$ represent a rooted weighted tree, where the leaf set $\vstarleaf$ represents distinct causes of death labeled as $c=1, \ldots, C$. Each cause is mapped to one and only one leaf in $\cT^\ast_{w^*}$. Note that $\bY$ maps every row of data $\Xb$ to a leaf in the tree for causes $\cT_{w^*}^*$; but this link is only observed for deaths occurred in the source domains $\{i:D_i\neq 0\}$. Similarities between causes are then characterized by between-cause distances in $\cT_{w^*}^*$. We then specify a logistic Gaussian diffusion prior:
\begin{align}
\theta_{jk}^{(c)} = {\sf expit}\left(\beta_{jk}^{(c)}\right),~~\beta_{jk}^{(c)} & = \sum_{u\in a(c)}\gamma_{jk}^{(u)}, c\in [C], \label{eq:cause_leaf}
\end{align}
with Gaussian increments over the edges leading to each leaf:
\begin{align}
\gamma_{jk}^{(u)} & \sim N(0,\tau^*_{\ell^*_u}w^*_u),\label{eq:cause_diffusion}
\end{align}
where $\ell^*_u$ maps node $u$ in the cause tree to one of $L^*$ level; this is to allow distinct diffusion variances for $L^*$ nonoverlapping blocks of a pre-specified partition of the nodes in the cause tree. Let $\btau^* = (\tau^*_1, \ldots, \tau^*_{L^*})^\transp$ be the vector of diffusion variances for the $L^*$ sets of nodes in the cause tree. Unlike in Section \ref{sec:domain_hieararchy}, we choose not to use node-specific spike-and-slab priors in the cause tree, which is equivalent to less aggressive shrinkage between causes and performs well in our simulation and validation studies.

\begin{remark}
Taken together, Sections \ref{sec:domain_hieararchy} and \ref{sec:cause_hierarchy} propose a prior for conditional distributions of the VA responses given any cause is a two-way tree-structured priors: i) the shrinkage among the domains in the columns is guided by a domain tree, and, ii) the shrinkage among the causes in the rows is guided by a cause tree. In particular, the shrinkage across causes is not domain-specific, but rather determined by information pooled across all domains. However, the shrinkage across domains is determined by a global-local structure, where we use $\tau_{\ell_u}$ as diffusion variance parameter for all causes (``global"; $\tau_{\ell_u}$ not indexed by $c$) and we use $\rho_{c\ell_u}$ to introduce cause-specific (``local"; $\rho_{c\ell_u}$ indexed by $c$) shrinkage of a leaf towards its parent ($\rho_{c\ell_u}$ closer to 0 or 1 for stronger or weaker shrinkage). 

% The global-local setup for shrinkage among the domains is introduced to allow strong shrinkage towards a parent for a domain $u=g$ with a small number of deaths of a certain cause $\rho_{c\ell_u}\approx 0$; for other domains with sufficient numbers of deaths due to the same cause, we can specify a uniform prior for $\rho_{c'\ell_{u}}$ owing to the additional flexibility. We did not introduce shrinkage among causes that allows local to each domain (i.e., $\tau^*_{\ell^*_u}$ not indexed by domain $g\in \{0\}\cup [G]$); this pools information across the domains to determine similarity between causes.
\end{remark}

% \zw{May21 update: Do not decouple $\ell_u$, just use more conservative - to encourage fusion.}

\subsection{Joint distribution} The joint distribution is fully specified by Equations (\ref{eq:csmf}-\ref{eq:nlcm_response}), (\ref{eq:logitSB}-\ref{eq::hyper_prior_with_level}), (\ref{eq:cmsf_dirichlet}), and  (\ref{eq:cause_leaf}-\ref{eq:cause_diffusion}). We collect the unobserved quantities by $\bGamma:=\{\bY^{\sf mis},\bpi^{(g)},\bZ,\balpha^{(c,u)},\gamma_{jk}^{(c)},\bs,\brho,c\in[C],g\in\{0\}\cup[G],j\in[J],k\in[K]\}$. We have the joint distribution of data $\cD$ and $\bGamma$ as follows:
\begin{align}
\label{eq:joint_all}
\begin{split}
\MoveEqLeft{ {\pr(\cD,\bGamma \mid \cT_w,\cT_{w^*}^*,\ba,\bb,\btau, \btau^*,\bd)}}  \\
=  &  \prod_{i=1}^{N}\prod_{g=0}^G \prod_{c=1}^C  \left(\pi_c^{(g)} \prod_{k=1}^K  \left\{\lambda_{k}^{(c,g)}\prod_{j=1}^J\sigma(X^*_{ij}\beta_{jk}^{(c)})\right\}^{\mathbf{1}\{Z_i=k,Y_i=c,D_i=g\}} \right)\\
& \times \prod_{c=1}^C\prod_{u\in \cV} \prod_{k=1}^{K-1} \frac{1}{\sqrt{2\pi\tau_{\ell_{u}}w_u}}\exp\left(-\frac{1}{2\tau_{\ell_{u}}w_u}\squared{\alpha^{(c,u)}_{k}}\right)\\
& \times \prod_{u\in \cV^*} \prod_{j=1}^J\prod_{k=1}^K \frac{1}{\sqrt{2\pi\tau^*_{\ell^*_u} w^*_u}}\exp\left(-\frac{1}{2\tau^*_{\ell^*_u}w^*_u}\squared{\gamma^{(u)}_{jk}}\right)\\
& \times \prod_{c=1}^{C}\prod_{u\in \cV} \rho_{c\ell_{u}}^{s_{cu}} (1-\rho_{c\ell_{u}})^{1-s_{cu}}\cdot  \prod_{g=0}^G {\sf Dirichlet}(\bpi^{(g)}; \bd^{(g)}) \times \prod_{c=1}^C\prod_{\ell=1}^{L} {\sf Beta}(\rho_{c\ell}; a_{c\ell},b_{c\ell}),
\end{split}
\end{align}
where Bernoulli likelihood components $\sigma(X^*_{ij}\beta_{jk}^{(c)}):= \{\theta_{jk}^{(c)}\}^{X_{ij}}\{1-\theta_{jk}^{(c)}\}^{1-X_{ij}}$ with $X^*_{ij} := 2X_{ij}-1$. Our primary quantity of interest is the CSMFs in the target domain $\bpi^{(0)}$ and the individual-specific cause-specific posterior probabilities $\PP(Y_i=c \mid D_i=0, \cD)$, $c\in[C]$.

The directed acyclic graph (DAG) in \ref{fig:doubletree_model_dag} in the Supplementary Materials shows the relationship between the observables, unknown quantities and hyper-parameters.

\section{Bayesian inference algorithms}
\label{sec:algorithm}

Calculating a posterior distribution often involves intractable high-dimensional integration over the unknowns in the model. Traditional sequential sampling approaches such as Markov chain Monte Carlo (MCMC) remains a widely used inferential tool based on approximate samples from the posterior distribution. They can be powerful in evaluating multidimensional integrals. However, they do not guarantee closed-form posterior distributions. {Variational inference (VI) is a popular alternative to MCMC for approximating the posterior distribution and has been widely used in machine learning and gaining interest in statistics \citep[e.g.,][]{blei2017variational,ormerod2010explaining}.} In particular, VI has also been used for fitting the classical LCMs \citep[e.g.,][]{grimmer2011introduction}. VI requires a user-specified family of distributions that can be expressed in tractable forms while being flexible enough to approximate the true posterior; the approximating distributions and their parameters are referred to as ``variational distributions" and ``variational parameters", respectively. VI algorithms find the best variational distribution that minimizes the Kullback-Leibler (KL) distance between the variational family and the true posterior distribution. VI has been widely applied in Gaussian \citep{carbonetto2012scalable,titsias2011spike} and binary likelihoods \citep[e.g.,][]{jaakkola2000bayesian,thomas2019estimating}. Also see \citet{blei2017variational} for a detailed review. We use VI because it is fast, bypasses infeasible analytic integration or data augmentation that is otherwise needed for MCMC under Dirac spike components and prior-likelihood non-conjugacy \citep{tuchler2008bayesian}, and enables data-driven selection of hyperparameters via approximate empirical Bayes (Step 3, \ref{secsupp::vi} in the Supplementary Materials).

We wish to obtain the marginal posterior distributions $\pr(\bpi^{(0)}\mid \cD)$ and $\pr(Z_i\mid \cD)$. We conduct posterior inference via variational inference. We assume the variational distributions can factorize as follows:
\begin{align}
    q(\bGamma) 
    =&  \prod_{g=0}^{G}q(\bpi^{(g)})\prod_{c=1}^C\prod_{u\in \cV} q(s_{cu},\balpha^{(c,u)})  \prod_{i:D_i=0} q(Y_i) \prod_{i=1}^{N} q(Z_{i}) \prod_{c=1}^C\prod_{\ell=1}^{L}q(\rho_{c\ell})\prod_{u\in\cV^*}\prod_{j,k}q(\gamma_{jk}^{(u)}).\label{eq:VIfamily}
\end{align}
% We update each factor in order while fixing the rest of factors by taking the expectation of a lower bound $H(\Xb^*,\bGamma)\leq \log \pr(\cD,\bGamma)$, the log-joint distribution of all the unknowns and observed data, with respect to the rest of factors in $q$. In particular, we lower bound the log-joint-distribution $\log \pr(\cD,\bGamma)$ by terms that are quadratic in $\alpha^{(c,u)}_{k}$ and $\gamma^{(u)}_{jk}$, which affords closed-form updates.

By the well-known equality, $\log\pr(\cD) = \cE(q)+KL(q||\pr(\bGamma \mid \cD))$. Because $\log \pr(\cD)$ is constant in $q$, minimizing the KL divergence between the variational family and the true posterior distribution is equivalent to maximizing $\cE(q)$, or ``evidence lower bound (ELBO)" which is defined by $\cE(q)  = \int q(\bGamma)\log \frac{\pr(\cD,\bGamma)}{q(\bGamma)} d \bGamma$ where $\bGamma$ collects all the unknowns.  We further bound $\cE(q)$ from below by bounding terms in $\pr(\cD,\bGamma)$ that involve sigmoid functions that create non-conjugacy issues under the Gaussian-distributed priors used in this paper hindering simple closed-form VI updates. In particular, following \citet{jaakkola2000bayesian}, we can bound Equation (\ref{eq:logitSB}) and $\sigma(X^*_{ij}\beta_{jk}\bracsup{c})$ from below respectively by
\begin{align}
\lambda_{k}\bracsup{c,g} \geq \{h(\eta_k^{(c,g)}; \phi_k^{(c,g)})\}^{\mathbf{1}\{k<K\}} \prod_{s<k} h(-\eta_s^{(c,g)}; \phi_s^{(c,g)}), g=0,1,\ldots,G,\textrm{~and~} \label{eq:bd1}
\end{align}
\begin{align}
 \sigma(X^*_{ij}\beta_{jk}\bracsup{c}) \geq h(X^*_{ij} \beta_{jk}\bracsup{c};\psi\bracsup{c}_{jk}),\label{eq:bd2}
\end{align}
where we have used the inequality
\begin{equation}
   h(x, \psi) := \sigma(\psi)\exp\{(x-\psi)/2-g(\psi)(x^2-\psi^2)\} \leq  \sigma(x), \label{eq::sigmoid:ineq}
\end{equation}
with $g(\psi) = \frac{1}{2\psi}[\sigma(\psi)-\frac{1}{2}]$ where $\psi$ is a tuning parameter. As a result, the right-hand-side terms in Equations (\ref{eq:bd1}) and (\ref{eq:bd2}) are quadratic in $\alpha^{(c,u)}_{k}$ and $\gamma^{(u)}_{jk}$, paving the way for closed-form VI updates. Also see \citet{durante2019conditionally} for a modern view of the technique as a bona fide variational algorithm with P{\'o}lya-Gamma augmentation. We now have a lower bound $\cE^\ast(q)$ of $\cE(q)$ which is defined as
\begin{align}
  \cE^*(q) & := \int q(\bGamma) \log\frac{H(\cD,\bGamma;\bpsi,\bphi)}{q(\bGamma)}\mathrm{d}\bGamma \leq \int q(\bGamma) \log\frac{\pr(\cD,\bGamma)}{q(\bGamma)}\mathrm{d}\bGamma = \cE(q),\label{eq:lower_bd_prDGamma}
\end{align}
where $H\leq \pr(\cD,\bGamma)$ is obtained by applying the lower bounds in Equations (\ref{eq:bd1}) and (\ref{eq:bd2}) to relevant terms in (\ref{eq:joint_all}) and has tuning parameters $\bpsi:=\{\psi_{jk}^{(c)},j\in[J],k\in[K],c\in[C]\}$ and $\bphi:=\{\phi^{(c,g)}_k,c\in[C],g\in\{0\}\cup[G],k\in[K-1]\}$; see \ref{secappend::logH} in the Supplementary Materials for the exact formula for calculation.

% Here the local variational parameters $\phi_k\bracsup{c,g}$  do not depend on the sign in front of $\eta_s\bracsup{c,g}$; And  $\psi_{jk}\bracsup{c}$ is not indexed by subjects hence not dependent upon $X^*_{ij}$ being $1$ or $-1$. This is because the optimal values of $\phi_s\bracsup{c,g}$ and $\psi_{jk}\bracsup{c}$ do not vary by the sign of the argument in $\sigma(\cdot)$. 

% $H = h^*(\Xb^{(g)},\bbeta,\{\bpi^{(g)}: g\neq 0\},\bpsi)\cdot h^{**}(\Xb^{(0)},\bbeta,\blambda^{(\cdot,0)},\bpi^{(0)})\pr(\bs,\balpha,\bs^*,\bgamma,\brho,\brho^*,\bpi)$.

The variational algorithm then finds the optimal variational distribution in the variational family that maximizes $\cE^\ast(q)$. In particular, we take the  logarithm of the lower bound $H$ of the joint probability density for data and unknowns with respect to a variational distribution $q$: $\EE_q [\log H]$. The algorithm updates each factor in order while holding the rest fixed. The update for the $j$-th factor in the variational distribution is $\EE_{q_{-j}}[\log H]$, where $\EE_{q_{-j}}$ means taking expectations with respect to $q$ over all but the variables in the $j$-th factor in $q$. The logarithmic of $H$ can be written as in \ref{secappend::logH} in the Supplementary Materials. A desirable property of $\log H$ is that integration of $\log H$ with respect to each factor of $q$ is in closed-form, which is a key ingredient of each VI update. The pseudo-code for the VI updates are provided in Algorithm \ref{algorithm::algo1}. See \ref{secsupp::vi} in the Supplementary Materials for the details of each update.

\newpage

\begin{algorithm2e}
{\footnotesize

\setstretch{0.8}
\caption{Pseudocode of Variational Bayes Algorithm}\label{algorithm::algo1}

\DontPrintSemicolon
\SetKwInput{Input}{Data}

\SetKwInput{Initialize}{Initialize}
\SetKwInput{Return}{Return}

\Input{
\begin{itemize}
    \item[($a$)] Multivariate binary data $\Xb$
	\item[($c$)] The domain ids $\bD$;
	\item[($b$)] The cause ids $Y_i$ for subject $i$ with $D_i\neq 0$;
	\item[($d$)] A weighted rooted tree for domains $\cT_w = (\cT=(\cV,E), w)$: leaves $\cG=\{0\}\cup[G]\subset \cV$, edge lengths $\bw=(w_u)_{u\in \cV}$;
	\item[($e$)] A weighted rooted tree for causes $\cT_w^*=(\cT^*=(\cV^*,E^*),w^*)$: leaves $[C]\subset \cV^*$;
\end{itemize}
}
\SetKwInput{Input}{Fixed Hyperparameters}

\Input{
\begin{itemize}
    \item[($a'$)] The number of classes $K\geq 2$; levels $\ell_{u}\in[L]$ for all nodes $u\in \cV$;
    levels $\ell^*_u\in[L^*]$ for all nodes $u\in \cV^*$;
	\item[($b'$)] Hyperparameters for the prior probability of $s_{cu} =1$: $(a_{c\ell},b_{c\ell})$, $c\in [C]$, $\ell\in [L]$; 
\end{itemize}
}

\Initialize{
\begin{itemize}
    \item[($a''$)] $t \longleftarrow 0$; Initialize $q_t(\bs,\balpha)$ and $q_t(\bgamma)$ \tcp*{(see Step 0 in  Appendix A1)}
    \item[($b''$)] Set an initial ELBO $\cE^\ast_0 \longleftarrow 0$
\end{itemize}
}

$t \longleftarrow 1$; $\cE^\ast_1 \longleftarrow \cE^\ast_0 +2\epsilon$

\While{$|\cE^\ast_t-\cE^\ast_{t-1}|>\epsilon$}{
    $q_t(\bs,\balpha,\bgamma) \longleftarrow q_{t-1}(\bs,\balpha,\bgamma)$
   
    $\bphi^{(t)} \longleftarrow \bphi^{(t-1)}$; $\bpsi^{(t)} \longleftarrow \bpsi^{(t-1)}$
    
    $\btau_1^{(t)} \longleftarrow \btau_1^{(t-1)}$; $\btau_2^{(t)} \longleftarrow \btau_2^{(t-1)}$
    
    \For{$c \in \cC$}{
        
          \For{$i \in [N]$}{
            \For{$k \in [K]$}{
                          if ($D_i==0$){
                $e^{(t)}_{ic} \longleftarrow \argmax_{e_{ic}}\cE^\ast_t(q)$ \tcp*{(See Step 1a in Appendix A1)}
           }
          if ($D_i \in \{1,\ldots,G\}$){
          
                    $r_{ik}^{(t)} \longleftarrow \argmax_{r_{ik}}\cE^\ast_t(q)$ \tcp*{(See Step 1b in Appendix A1)}
            }
          }
        }
    }
    
     \For{$g=0,1,\ldots, G$}{
      $q_{t}(\bpi^{(g)}) \longleftarrow \argmax_{q_t(\bpi^{(g)})} \cE^*(q)$ \tcp*{(See Step 1c in Appendix A1)}
     }
    
    \For{$c \in \cC$}{
        \For{$u \in \cV$}{
            $q_t(s_{cu}, \balpha^{(c,u)}) \longleftarrow \argmax_{q_t(s_{cu}, \balpha^{(c,u)})}\cE^\ast_t(q)$ \tcp*{(see Step 1d in Appendix A1)}
        }
    }
    
    \For{$u \in \cV^*$}{
        $q_t(\bgamma^{(u)}) \longleftarrow \argmax_{q_t(\bgamma^{(u)})}\cE^\ast_t(q)$ \tcp*{(see Step 1e in Appendix A1)}
    }
    
    \For{$c \in \cC$}{
        \For{$\ell \in [L]$}{
            $q_t(\rho_{c\ell}) \longleftarrow \argmax_{q_t(\rho_{c\ell})}\cE^\ast_t(q)$ \tcp*{(see Step 1f in Appendix A1)}
        }
    }
    
%    \For{$\ell \in [L^*]$}{
%        $q_t(\rho^*_\ell) \longleftarrow \argmax_{q_t(\rho^*_\ell)}\cE^\ast_t(q)$ \tcp*{(see Step 1c in Appendix A1)}
%    }
    
        \For{$k \in [K]$}{
        \tcp*{update local variational parameters for tighter lower bounds}
          \For{$c \in \cC$}{
              \For{$g \in \cG$}{
                          $\phi^{(c,g),(t)}_k \longleftarrow \argmax_{\phi^{(c,g)}_k}\cE^\ast_t(q)$ 
                          }
                     
                \For{$j \in [J]$}{
                    $\psi^{(c),(t)}_{jk} \longleftarrow \argmax_{\psi^{(c)}_{jk}}\cE^\ast_t(q)$ \tcp*{(see Step 2 in Appendix A1)}
                }
             }
        }

    \If{$t~\text{mod}~d=0$}{

              \For{$\ell \in [L]$}{
                      $\tau^{(t)}_{\ell} \longleftarrow\argmax_{\tau_{l}}\cE^\ast_t(q)$ %\tcp*{(see Step 3 in Appendix \ref{secsupp::vi})}
                      }
           
        \For{$\ell \in [L^*]$}{
            $\tau^{*,(t)}_{\ell} \longleftarrow \argmax_{\tau^*_{\ell}}\cE^\ast_t(q)$ \tcp*{(see Step 3 in Appendix A1)}    
        }
    }
 $\cE^\ast_t \longleftarrow ELBO(q_t)$ \tcp*{(see Step 4 in Appendix A1)}
 
 $t \longleftarrow t+1$
 }% end while

\Return{$q_{t-1}(\bs,\balpha)$, $q_{t-1}(\bgamma)$, $\{q_{t-1}(Y_i),D_i=0\}$, $q_{t-1}(\brho)$, $\{\cE^\ast_1, \ldots, \cE^\ast_{t-1}\}$}

}%end change size

\end{algorithm2e}

\paragraph{Cause-specific domain dissimilarity measure} The framework motivates the following the estimated cophenetic distance \citep[e.g.,][]{sneath1973numerical} between the target domain and each of the source domains. In particular,  the dissimilarity measure is $dist_{\cT_{w'}}({\sf target=0},{\sf source}=g; {\sf cause}= c)$, $g=1,\ldots, G$, $c=1, \ldots, C$, where $w'(P)$ is defined as the sum of all the modified edge weights along the path $P$ connecting the target domain and source domain $g$ in $\cT_{w'}$, and the modified weight of an edge $(pa(u)\rightarrow u)$ is $w'_u = {p}_{cu}\cdot w_u$ for node $u\in \cV$ in the domain tree (see Equation (\ref{appeq:vi_prob1}) of VI updates in \ref{secsupp::vi} in the Supplementary Materials for definition of $p_{cu}$ (on a logit scale) as the variational approximation to $\PP(s_{cu}=1\mid \cD)$). 

\paragraph{Choice of $K$} We follow \cite{bishop2006pattern} and use criterion $\cE_K^\ast(q)+\log(K!)$ where $\cE_K^\ast(q)$ is the lower bound of log marginal data likelihood for a $K$-class model and the correction term is to make different models comparable \cite[e.g.,][Section 5.2]{grimmer2011introduction}.

\paragraph{Software} A free and publicly available \verb"R" package that implements the VI algorithm for scalable approximate posterior inference is freely available at \url{https://github.com/zhenkewu/doubletree}. The package is designed to work under all possible patterns of observed and missing causes of death: ({\sf Scenario i}) $Y_i$ is missing in a single domain (say $g$): {\sf i-1}) none has confirmed causes in domain $g$; {\sf i-2}) there exists at least one observed $Y_i$ in domain $g$; {(\sf Scenario ii}) $Y_i$ is missing in $M\geq 2$ domains; {(\sf Scenario iii)} no missing $Y_i$ in any domain. This paper has focused on {\sf Scenario (i-1)} for simpler exposition.

\section{Simulation Studies}
\label{sec:simulation}

We conduct two sets of simulation studies to evaluate the operating characteristics of the proposed method and demonstrate its better capability of estimating the target-domain CSMFs and assigning individual-level causes of death relative to a few alternatives with ad hoc specifications of information pooling across the domains. In the first set of simulations, we simulate data based on true parameters values under NLCM. In the second set of simulations, we use a validation data set and selectively mask a subset of deaths' true causes in a synthetically constructed target domain and then apply NLCM. The simulation designs, performance metrics, and results are detailed below.

\paragraph{Performance Metrics} 
First, we assess the overall accuracy by the so-called ``CSMF accuracy'' \citep{murray2011robust}, widely used as the metric to compare the estimated CSMF vector against the truth. The CSMF accuracy metric measures the $L_1$-distance between the estimated and true vectors of CSMFs, and is normalized to range between $0$ (worst) and $1$ (best). It is defined as \({\sf CSMF}_{\sf acc}(\hat\pi^{(0)}) = 1 - \frac{\sum_{c=1}^C |\hat\pi^{(0)}_c - \pi^{(0)}_c|}{2(1 - \min_{c} \pi^{(0)}_c)},\) where $\pi^{(0)}_c$ is the true CSMF for cause $c$ that we set in simulation design (or calculated by the empirical distribution in the synthetic target domain data in Simulation II below). This formulation is also known as the normalized absolute error in the quantification learning literature \citep{gonzalez2017review}. Finally, for the accuracy of COD classifications, we will use top cause accuracy: the fraction of deaths with the true CODs in the top predicted causes.

\subsection{Simulation I} 

\paragraph{Design} 

We simulated $R=200$ independent replicate data sets for different total sample sizes $(N=1000,4000)$. To illustrate, we use a domain tree $\cT_w$ shown in Figure \ref{fig::simI_tree_setup} with equal edge weights and true domain leaf groups; $\cT_w$ has $\pleaf=6$ leaves and $3$ domain leaf groups. The leaf ``0" is set to be the target domain leaf. For each $N$, we set each domain's sample size to be approximately $N/\pleaf$ for $g =0,1,\ldots,G$ (with rounding where needed) to investigate balanced leaves and set the sample size in domain $g$ to be approximately $\frac{1}{5}N/\pleaf$ or $\frac{4}{5}N/\pleaf$ with equal chances for mimicking unbalanced observations across the domain leaves. Within domain $g$, we further assign deaths into $C$ causes by independently sampling from categorical distributions with CSMFs $\bpi_g$, $g=0,1,\ldots, G$. We then simulated multivariate binary response data for different dimensions $J=20, 60$, for $K=2$ classes according to Equations (\ref{eq:nlcm_class}) and (\ref{eq:nlcm_response}). We considered $C=3$ causes. Two different sets of $\{\bTheta^{(c)}\}$ were considered; see \ref{secappendix::simulation_details} in the Supplementary Materials for more details of the true parameter values and model setup.

For each simulated data set, we fitted the proposed model, based on which we compute the approximate posterior mean of $\bpi^{(0)}$ obtained via its optimal variational distribution. In addition, we also compared against a few NLCM-based approaches but with suboptimal, ad hoc specifications of information pooling between the domains (to different degrees of cross-domain shrinkage). Figure \ref{fig::simI_tree_setup} shows the domain groupings used in these comparisons. In summary, we fit 1) the proposed method:``{\sf Domain Adaptive}"; 2) ``{\sf True Domain Grouping}": for any cause $c$, assume identical $\pr(\bX_i \mid Y_i=c,D_i=g)$ for domains in a group (4 groups in the simulation truth). To do so, for each $c$, we fix $s_{cu}$ to 1s or 0s in a way that results in the true domain grouping; 3) ``{\sf Complete Pooling}": completely ignore the external domain tree information during estimation and also ignore the sample-to-domain mappings $\bD$. By doing so, we assume $\pr(\bX_i \mid Y_i=c,D_i=g)$ remains the same between the domains; 4) ``{\sf Ad hoc Domain Grouping}": use a manual domain grouping that is finer than the true domain grouping; 5) ``{\sf No Domain Grouping}": same as 3) except $\bD$ is used during estimation so that data are recognized to have come from different domains.

% Third, we compared the true and the estimated leaf groupings via adjusted Rand Index \citep[ARI,][]{hubert1985comparing}. ARI is a chance-corrected index that takes value between $-1$ and $1$ with values closer to $1$ indicating better agreement. Fourth, we estimated the coverage probability of the approximate $95\%$ CrIs. For each true group $g$, we compute the frequency of the approximate $95\%$ CrI (computed along with $\hat{\bpi}^{\sf dgrp}_v$) containing the truth, conditional on the event that an estimated partition of the leaf nodes includes $g$.

\paragraph{Results} 

% We compare five methods. The first one is the proposed method, where we perform tree-structured shrinkage over the domains. The second method is to set $s_{cu}$ to particular values so that we form the true grouping of domains. The third method is to ignore the domain tree structure and pool data from all the domains under the assumption that the conditional distributions of the multivariate binary responses given the cause remain the same. The fourth method is based on ad hoc domain grouping in a way different from the true domain grouping; in our simulation, the ad hoc grouping is more granular than the truth, but contains the true grouping of {\sf AP, UP} - all other groups are of size one. The final method does not encourage similarity among any subset of domains, but does share the class-specific response probability parameters across domains; with domain-specific class mixing weights, the conditional distributions of the multivariate responses given a cause may differ between domains. These five methods are referred to as NLCM-DT, True Domain Grouping, Pooled, Ad Hoc Domain Grouping, and Target-domain only respectively in the resulting figures. 
% \ic{Just a suggestion since this is quite a substantial paper and it might make the readers' lives a little easier by spelling out which methods are which in the results. :) }

Figure \ref{fig::simI_csmf_acc} compares the methods in terms of CSMF accuracy. The proposed NLCM Domain Adaptive method adaptively learns the the domain groupings and produced the CSMF estimate for the target domain with the highest accuracy. The accuracy is comparable to the ones obtained under the true domain grouping. The method with complete pooling between domains generally performs the worst; this should not be surprising given the simulation truth is to mimic situations where $\pr(X_i \mid Y_i,D_i=g)$ differs by domain. NLCMs with ad hoc domain grouping and no grouping produced more accurate estimates than the one with complete pooling between the domains. However, because both methods are based on domain groupings that are finer than the true domain grouping, they do not fully use similar domains to improve the accuracy of estimating the conditional distributions of the VA responses given a cause, resulting in sizable losses in CSMF accuracy. Similar relative patterns are also clear when RMSEs are compared (see Figure \ref{fig::simI_RMSE}). In addition, as expected the conditional dependence modeled by the NLCMs for each cause improved the individual-level classification performance (top cause accuracy) relative to methods that ignored conditional dependence (results not shown here).

\begin{figure}[htp]
\captionsetup{width=0.9\textwidth}
\centering
\addtocounter{figure}{0} 
{\subfigure[Simulation I: domain tree and different domain groupings used in comparison.]{
\includegraphics[width=.45\linewidth]{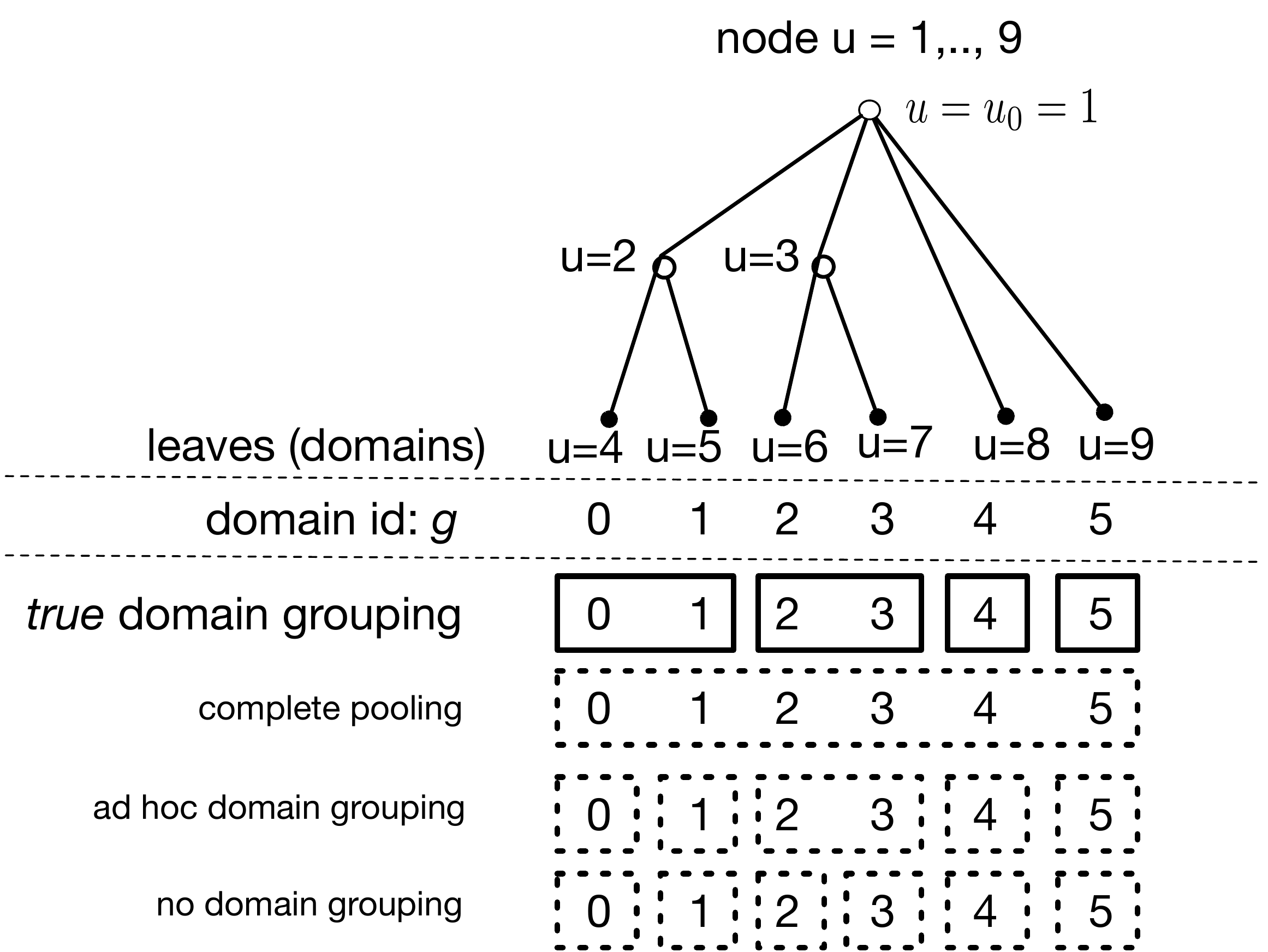}
\label{fig::simI_tree_setup}
}
}
{\subfigure[Simulation I: CSMF accuracy comparison.]{
\includegraphics[width=.9\textwidth]{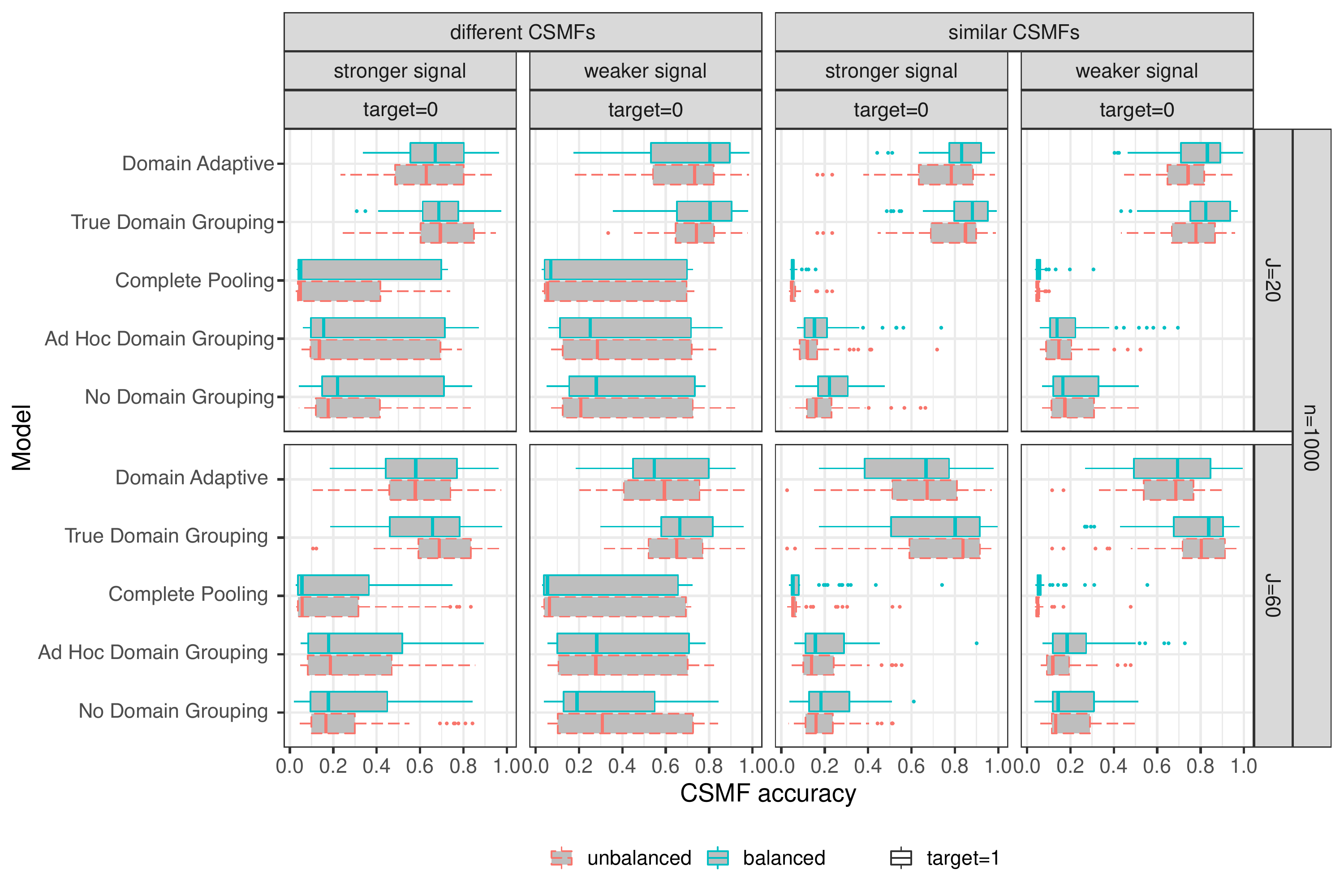}
\label{fig::simI_csmf_acc}
}
}
\addtocounter{figure}{0} 
\caption{Simulation I: domain tree setup and results.}
\label{fig:simI}
\end{figure}

\subsection{PHMRC VA Data: Background and Description} 
\label{sec:databackground}
In Section \ref{secapp:simulationII} and Section \ref{sec:dataapp}, we will use the Population Health Metrics Research Consortium (PHMRC) VA validation data with physician-coded causes of death to evaluate the performance of the proposed method. We first review aspects of the data set pertinent to our study. 

%The data curation script can be accessed at \url{https://github.com/richardli/multiVA/blob/master/codes/phmrc-preprocess.R}.

PHMRC VA validation data collection was implemented in six sites in four countries: Andhra Pradesh (``AP"), India; Uttar Pradesh (``UP"), India; Dar es Salaam (``Dar"), Tanzania; Pemba Island (``Pemba"), Tanzania; Bohol, Philippines; Mexico City (``Mexico"), Mexico. The goal is to create a high quality validation data set from different populations to evaluate comparative method performance and make recommendations for future VA implementations.  See \citet{murray2011population} for a more complete description of the PHMRC VA validation data. The data set with 7,841 adult deaths and 168 symptoms that we use here is further based on preprocessing in \citet{mccormick2016probabilistic}.

% A cause list was constructed based on the WHO Global Burden of Disease estimates of the leading causes of death, potential to identify unique signs and symptoms, and the likely existence of sufficient medical technology to ascertain gold standard cases. Blinded verbal autopsies were collected on all gold-standard deaths. The cause lists were also designed so that they were mutually exclusive and collectively exhaustive. The target cause list for adults, children, and neonates included 53 gold-standard causes. 

Figure \ref{fig:twotrees_deathcounts} shows the domain hierarchy via a rooted tree with six leaves at the top margin. The domain hierarchy uses country membership information to form the leaves and internal nodes. Unless otherwise stated, we set all edge weights to be one for illustration. External domain-level information that is highly associated with CSMFs may also be incorporated to form the domain hierarchy, e.g., via hierarchical clustering of a variety of domain-level information that may alter symptom-cause relationships, such as time periods, level of VA interviewer training, and differential availability of treatments that mitigate a subset of symptoms interviewed in VA. As a secondary feature of the proposed method, the left margin of Figure \ref{fig:twotrees_deathcounts} shows the cause hierarchy with $35$ leaves along with coarser aggregated cause definitions.

Among the 168 symptoms, 63 have a missing rate of higher than $1\%$, of which 37 has a missing rate higher than $5\%$. The highest missing rate is $96.6\%$ for: ``Was there pain in the upper belly?"; the next highest missing rate is $92\%$, for four questions related to where rash was located if present: ``Trunk?",``Extremities?","Everywhere?", ``Other locations?" In addition, the numbers of missing symptoms for a subject are between 1 and 76, with a median of 22.  Missing data in VA in the form of ``Don't Know" or ``Choose not to answer" appear for various reasons. Missing data have been considered by \citet{kunihama2020bayesian} under the assumption of missing at random. Absent additional information regarding individual-specific missing data mechanism and given the lack of likely alternative sensitivity assumptions about missing data, we will assume missing at random in this paper when conducting model estimation and method comparisons. This sets the stage for fair relative comparisons with other existing methods that either assume missing completely at random or missing at random. In our proposed model, because given a cause and a class, symptoms are assumed mutually independent, when calculating the cause-specific likelihood for an individual in class $k$ with a subset of missing symptoms, we simply do products of Bernoulli likelihoods only over symptoms with non-missing information; see the implementation in Steps 1a and 1e in \ref{secsupp::vi} during variational updates.

\begin{figure}[htp]
\captionsetup{width=\linewidth}
\centering
\addtocounter{figure}{0} 
\raisebox{0in}{
%\subfigure[]{
 \includegraphics[width=.95\linewidth]
{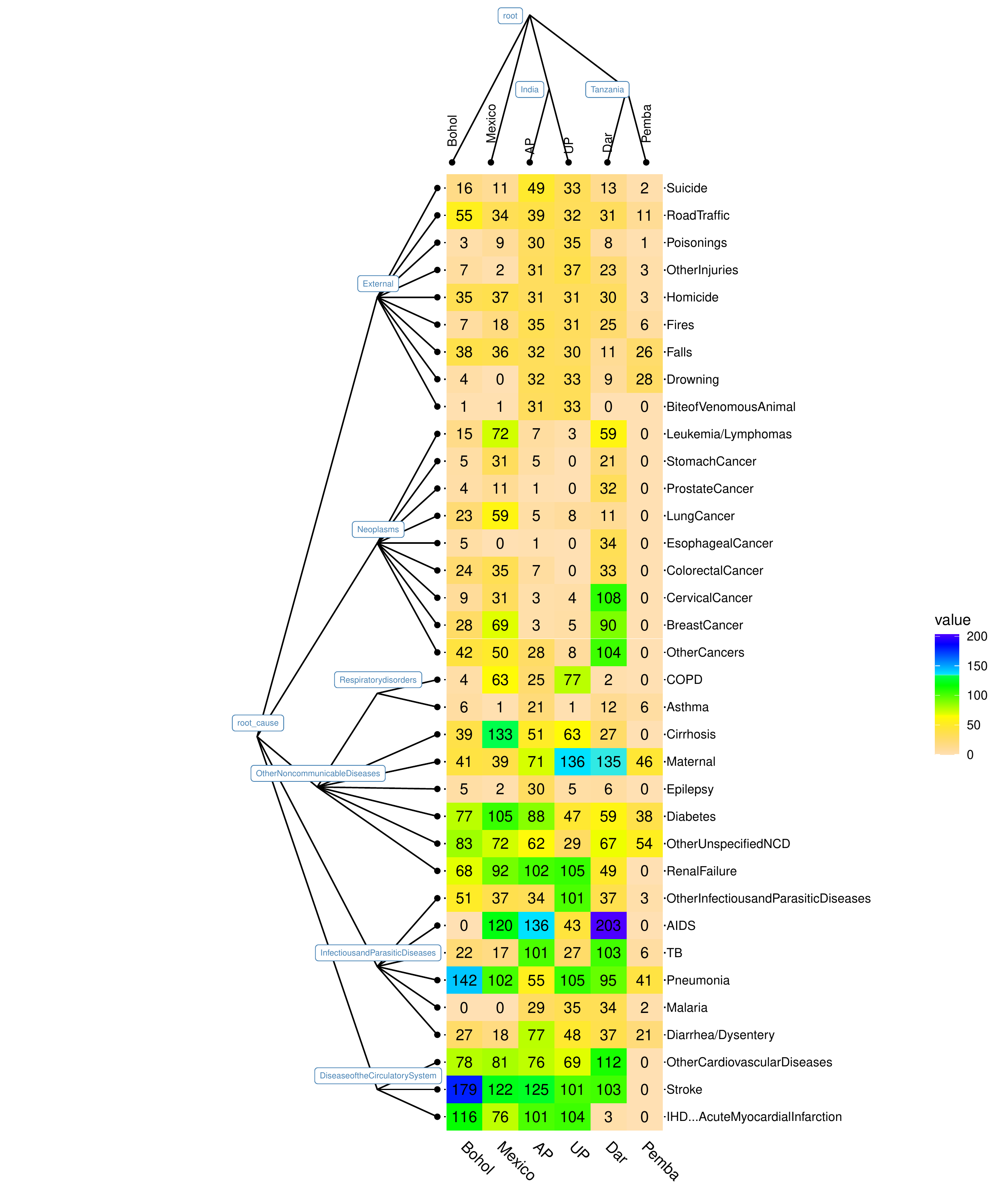}
%\label{fig:twotrees_heatmap}
%}
}
%\raisebox{-.5in}{
%\subfigure[Scenario 2 with a larger tree]{
% \includegraphics[width=.47\linewidth]
%{real_tree.pdf}
%\label{fig:simu_tree2}
%}
%}
%}

\addtocounter{figure}{0} 
\caption{Death counts by cause and site for $N=7,841$ deaths and $J=168$ across all six sites in the PHMRC data set. The exact death counts are shown in corresponding cells. Shown on the left and top margins are the cause and domain hierarchies assumed in the data analysis. We will mask the causes-of-death in one site during method testing so the site with masked causes is the target domain and the rest sites are source domains. The domains closer in the domain hierarchy (top margin) are \textit{a priori} more likely fused to have the same vector of class mixing weights. The proposed method has a secondary feature that can incorporate a cause hierarchy (left margin) with 35 causes on the leaves and six aggregated causes represented by internal nodes.}
\label{fig:twotrees_deathcounts}
\end{figure}

\subsection{Simulation II: Semi-Synthetic}
Here we use the PHMRC VA validation data to evaluate the performance of the proposed method. Because validation data contains gold-standard causes of death, we split the original data set into source and target domain data sets and mask the causes of the deaths allocated to the target domain. Because we use the real PHMRC data while masking a subset of causes of death, we refer to this simulation as ``semi-synthetic". See  \ref{secapp:simulationII} in the Supplementary Materials for the details about the design and results showing that domain adaptive estimation provides more accurate CSMF estimates and top cause COD assignment.

\section{Domain Adaptation across Actual PHMRC Sites}
\label{sec:dataapp}

% The selection of the target cause list was based on consideration of the WHO estimates of the leading causes of death in the developing world in each age group, those causes for which verbal autopsy might be able to function adequately because unique signs and symptoms could potentially be collected in an interview, and the potential to find, in the six sites, deaths with sufficient laboratory, medical imaging, and pathological detail in order that a gold standard cause of death assignment could be made. 

%  We will show that under this setup, the method works well by accurately learning the conditional distributions despite differences in the CSMFs.

% We will compare a few methods: 1) NLCM-DT, 2) pooled analysis that assumes identical mixing weights between the domains, 3) NLCM based on ad hoc domain grouping, 4) target-domain-only analysis (but with shared response probabilities across the domains for each cause); we did not have the NLCM-DT comparator because we do not know the true domain grouping in the PHMRC data.

This scenario uses the actual PHMRC domain designation: one site as the target domain with the gold-standard cause-of-death labels masked, the rest five sites as source domains with observed gold-standard causes-of-death. Table \ref{table::simulationIIc} shows the CMSF accuracy when each of the six PHMRC study sites is treated as the target domain iteratively. The method selected two-class models. The domain adaptive approach achieved better accuracies in CSMF estimation and slight improvements on the top-cause classification accuracies. The somewhat low accuracies using PHMRC data are well known, motivating new ongoing validation data collection by our substantive collaborators based on which we will further test our method. To illustrate how to interpret results from the domain adaptive method, we pick {\sf AP} as the target domain. In addition, we pick six causes (out of $35$ causes used during model fittings) to illustrate the results of cause-specific shrinkage. The causes are selected based on the different levels of shrinkage: near-complete pooling (cause ``Drowning"), and no substantial shrinkage between the domains (cause AIDS, Stroke, Renal Failure, Tuberculosis (TB), Inflammatory Heart Disease (IHD)-Acute Myocardial Infarction (MI)).  

Using {\sf AP} as an example target domain, Figure \ref{fig:PHMRC_lcm} shows two-class LCM results for some pairs of (cause,domain). For each of the six causes shown in Figure \ref{fig:phmrc_class_weights}, the relative importance of the two classes varies greatly by domain, indicating substantive differences in $\pr(\bX_i \mid Y_i=c,D_i=g)$ between the domains. This is evident from the patterns of the class mixing weights for ``TB" and ``Renal Failure". Interestingly, this is not uniformly true for all causes, mostly notably for ``Drowning", which placed large weights on the first class regardless of domain. A plausible explanation for this empirical finding is that symptoms are easily recognizable and highly specific if a death is caused by drowning.  This analysis also empirically confirms that  the conditional distributions of VA responses given each cause may vary by cause. 

In addition, we also calculate the estimated cophenetic distance \citep[e.g.,][]{sneath1973numerical} between the target domain and each of the source domains (see Section \ref{sec:algorithm}). Figure \ref{fig:cophenetic} shows for each of 35 causes (rows), five estimated distances with smaller values representing higher similarity between the target domain ({\sf AP}) and each of the five source domains shown on the x-axis. We make a few interesting empirical observations. First, {\sf UP} is estimated to be most similar to the target domain for almost all causes, which is perhaps unsurprising given {\sf AP} and {\sf UP} are two states in India. Second, the joint pattern of similarity between {\sf AP} and five source domains differ by cause. For example, for causes like ``Drowning", ``Malaria", ``Asthma", the dissimilar measures are all estimated to be small, indicating {\sf AP} is uniformly similar to all other source domains in terms of symptom-cause conditional distributions. This uniformity in the dissimilarity measure may be explained by reporting of specific symptoms that vary little by domain. On the other hand, for causes with complex etiologies and manifest symptoms such as ``Renal Failure", the target domain {\sf AP} is estimated to be most similar to {\sf UP} and least so to {\sf Bohol}.

\begin{figure}[htp]
\captionsetup{width=0.9\linewidth}
\centering
\addtocounter{figure}{0} 
{\subfigure[Class-specific response probabilities based on a $K=2$ class model (top 5 causes in AP and Drowning; top 20 symptoms with the highest estimated marginal probabilities).]{
\includegraphics[width=.9\linewidth]{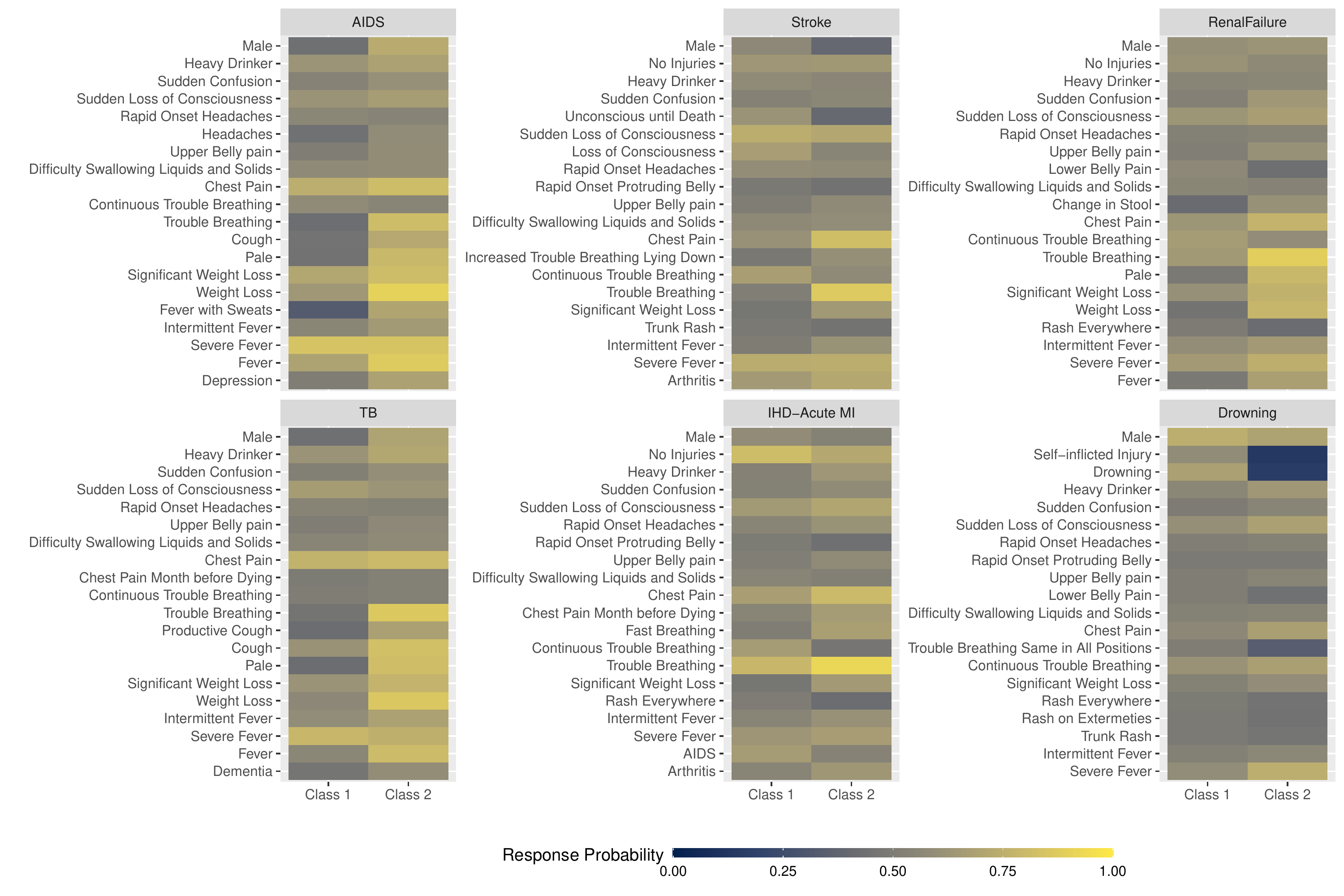}
\label{fig:phmrc_class_response_profiles}
}
}
{\subfigure[Variation of class-mixing weights between domains; six sets of weights are shown for six causes of deaths (the model uses 35 causes).]{
\includegraphics[width=.43\linewidth]{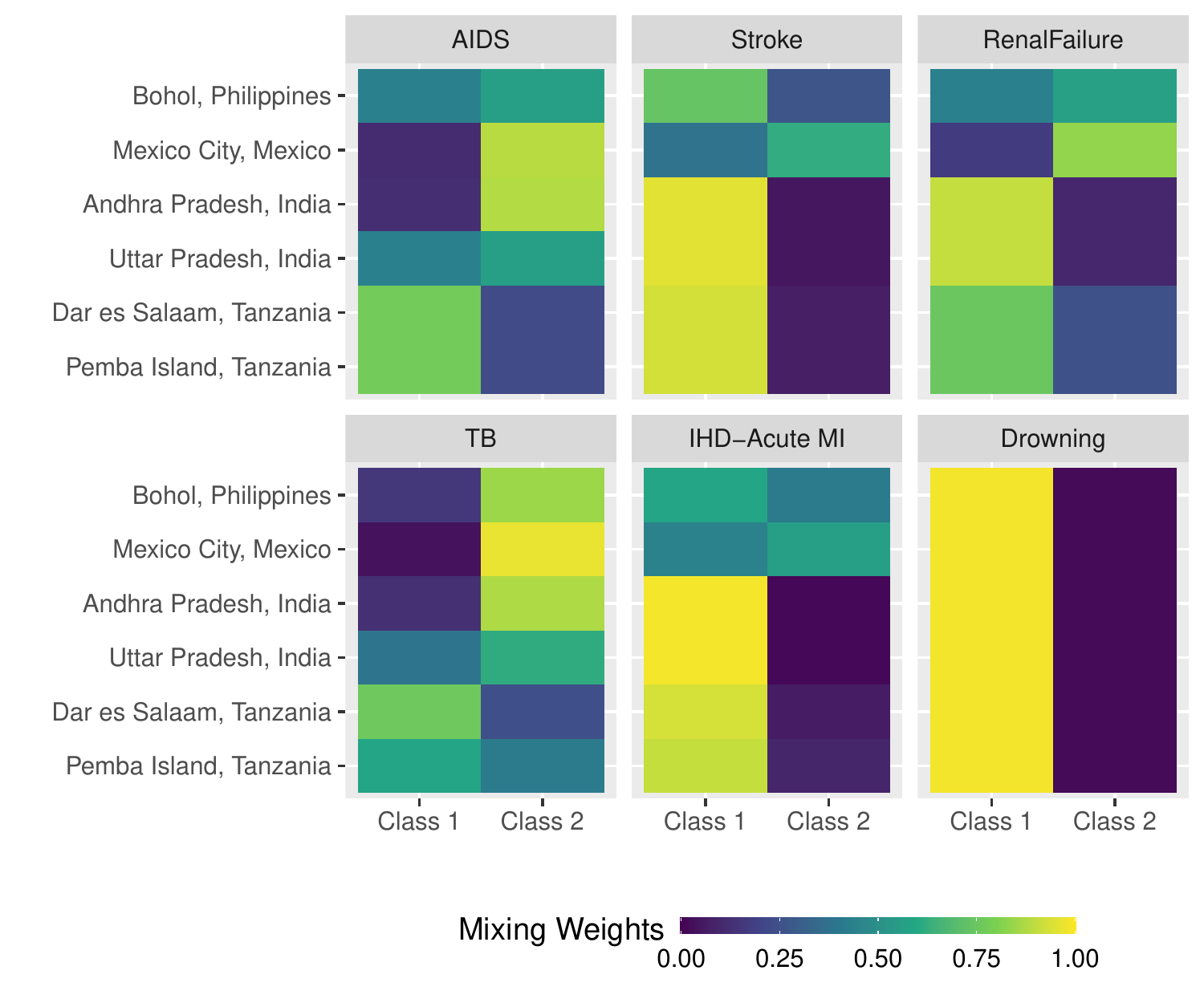}
\label{fig:phmrc_class_weights}
}
}\hspace{0.3cm}% 
{\subfigure[Estimated cause-specific cophenetic distances between AP (target) and each of the five source domains; 35 rows representing 35 causes used during model fitting.]{
\includegraphics[width=.43\linewidth]{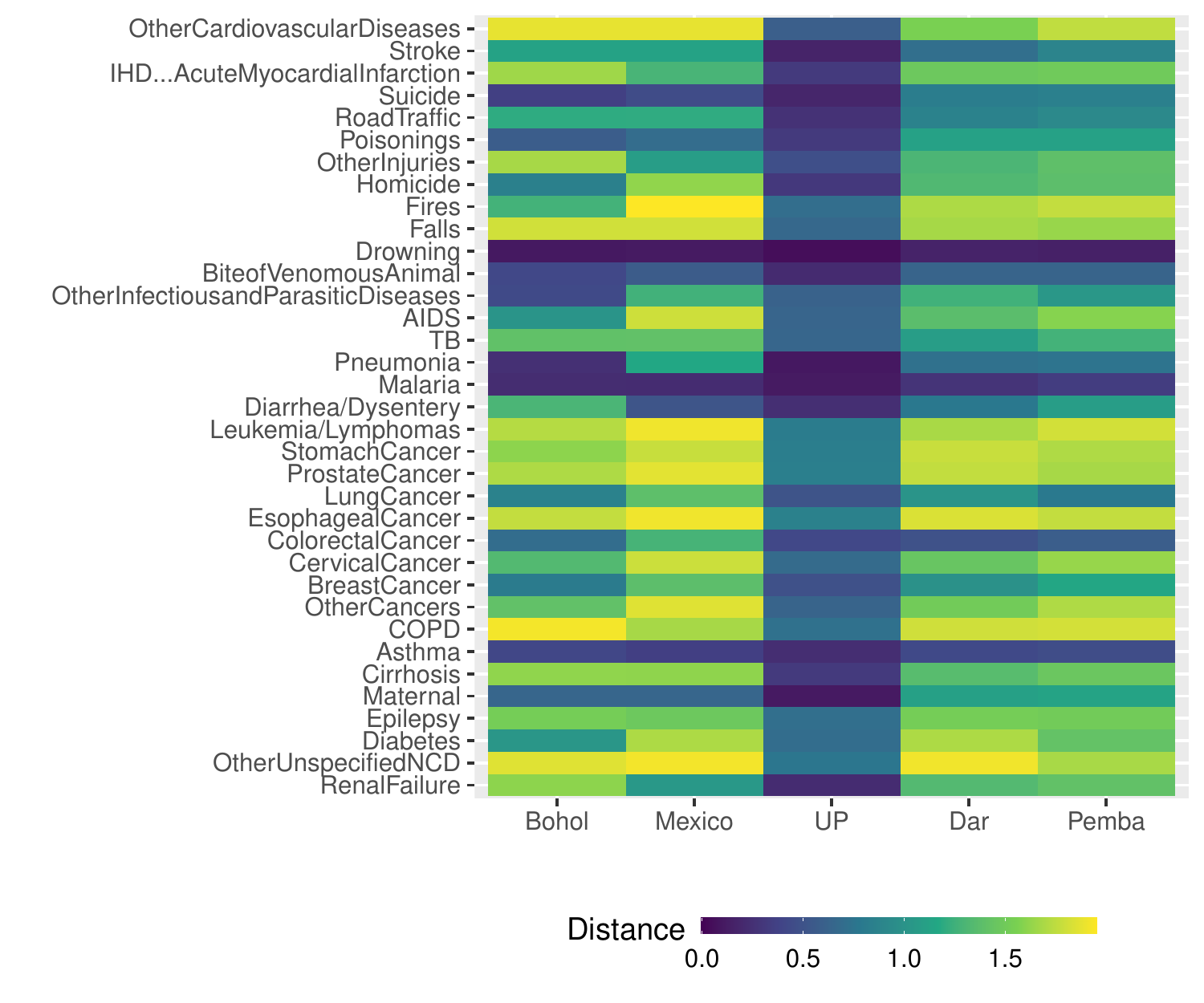}
\label{fig:cophenetic}
}
}
\addtocounter{figure}{0} 
\caption{PHMRC results based on a $K=2$ class nested LCM; for illustration, model results using {\sf AP} as the target domain are shown here.}
\label{fig:PHMRC_lcm}
\end{figure}

\begin{table}[htp]
\centering \captionsetup{width=0.7\textwidth}
\caption{CSMF accuracy when each of the six PHMRC study sites is treated as the target domain iteratively.}
\begin{tabular}{lllllll}
\hline\hline
Method            & Bohol & Mexico & AP   & UP   & Dar  & Pemba \\\hline 
                       & \multicolumn{6}{c}{\textit{CSMF Accuracy}}\\ 
Domain Adaptive               &  0.68  & 0.72   & 0.70 & 0.66 & 0.66 & 0.63  \\
% NLCM-DT                & 168 & 3 & 0.63  & 0.64   & 0.67 & 0.61 & 0.54 & 0.63  \\
Complete Pooling                 & 0.67  & 0.67    & 0.67 & 0.61 & 0.61 & 0.58  \\
% Pooled                 & 168 & 3 & 0.65  & 0.56   & 0.67 & 0.57 & 0.53 & 0.62  \\
Ad Hoc Domain Grouping & 0.67  & 0.68   & 0.68 & 0.61 & 0.65 & 0.62  \\
% Ad Hoc Domain Grouping & 168 & 3 & 0.65  & 0.58   & 0.68 & 0.55 & 0.55 & 0.58  \\
No Domain Grouping     &  0.67  & 0.64   & 0.66 & 0.62 & 0.63 & 0.60   \\
% Target-domain only     & 168 & 3 & 0.65  & 0.64   & 0.66 & 0.6  & 0.54 & 0.63 \\ 
\hline
                       & \multicolumn{6}{c}{\textit{Top Cause Accuracy}}\\ 
Domain Adaptive                &  0.32 & 0.30  & 0.34 & 0.36 & 0.34 & 0.41 \\
% NLCM-DT                & 168 & 3 & 0.3  & 0.27 & 0.36 & 0.29 & 0.29 & 0.41 \\
Complete Pooling                  & 0.32 & 0.29 & 0.35 & 0.36 & 0.33 & 0.39 \\
% Pooled                 & 168 & 3 & 0.3  & 0.27 & 0.38 & 0.29 & 0.29 & 0.39 \\
Ad Hoc Domain Grouping & 0.32 & 0.30  & 0.36 & 0.35 & 0.33 & 0.38 \\
% Ad Hoc Domain Grouping & 168 & 3 & 0.30  & 0.27 & 0.36 & 0.28 & 0.3  & 0.38 \\
No Domain Grouping     &  0.32 & 0.30  & 0.34 & 0.35 & 0.33 & 0.40  \\
% Target-domain only     & 168 & 3 & 0.3  & 0.27 & 0.35 & 0.3  & 0.29 & 0.39 \\
\hline\hline
\end{tabular}
\label{table::simulationIIc}
\end{table}

\section{Discussion}
\label{sec:discussion}

\paragraph{Summary} In this paper, we presented a hierarchical Bayesian approach to use individual-level multivariate binary responses obtained from a target domain in the absence of any gold-standard categorical labels (``causes" in VA) for the estimation of the target population fractions of the label categories (``CSMFs" in VA). This is made possible by using individual-level data from multiple source domains where additional gold-standard cause labels are available. The data from multiple domains are integrated following a data-driven tree-structured shrinkage approach, so that for each cause, domains that have similar conditional distributions of the responses given the cause are encouraged to be pooled to improve estimation. We achieved this goal via a logistic stick-breaking Gaussian diffusion process on the mixing weights along a pre-specified domain hierarchy. In addition, an analyst may use another hierarchy applied to causes to regularize the parameter estimation when characterizing the conditional distributions of the responses given a cause. Simulation studies show that the proposed method produces more accurate estimation of CSMFs than methods that either ignore between-domain differences by complete pooling of data across the domains, ad hoc specification of domain grouping, or no domain grouping at all; the proposed method performs similarly to an oracle method with the true domain groupings. Although this paper focuses on the more challenging case where no cause label is observed in the target domain for simpler exposition, the model readily generalizes to using gold-standard cause labels for a subset of deaths in the target domain. The software accompanying this paper has implemented such extensions.  

\paragraph{Limitations} Practical limitations of the proposed approach may exist. First, the hierarchy for the domain is pre-specified based on external domain-level information (geo-locations of the study sites  in this paper) and are not estimated from the VA data themselves. Methods that specify a prior over the space of domain hierarchies for posterior inference may be fruitful at extra computational costs \citep[e.g.,][]{knowles2014pitman}. Second, deviation from missing-at-random assumption for the VA questionnaire responses may impact the model results and performance during domain adaptations. This issue remains challenging and to be explored in collaboration with VA substantive experts to identify common and major reasons leading to missing data. Third, the cause list used in this paper is chosen based on clear clinical meanings of distinct etiologic implications and also for methodological illustration. Certain causes of death, such as COVID-19 related deaths, may quickly emerge as prominent causes in some populations that necessitates updated cause lists. Fourth, PHMRC data is currently the only validation data set for evaluation; future VA gold standard data sets will be available for additional validation. Fifth, additional unstructured narrative texts are available, and may improve the capacity of the present approach by augmenting the VA symptoms with absence or presence of derived text features (e.g., cause-discriminative words). Finally, the current results are agnostic to additional information about symptom-cause relationships. Estimation accuracy may be further improved by incorporating these information \citep[e.g.,][]{mccormick2016probabilistic,schifeling2016incorporating}.

% Finally, the present approach provides a similarity measure for each of the source domains by the spike probability weighted length of the path between a source domain and the target domain in the hierarchy. 

\paragraph{Future directions} There are a few statistical extensions that may further improve the utility of the proposed method. First, additional individual-level covariates, such as age, pregnancy status, and seasonality, that may explain variation in the conditional distribution of the VA questionnaire responses given a cause $\pr(\bX_i\mid Y_i,D_i)$. Our framework readily incorporates discrete individual-level covariates via concatenation with the vector of VA questionnaire responses. For continuous individual-level covariates, \citet{moran2021bayesian} illustrated an approach based on a different framework of factor models, which however is not suited to dealing with a domain absent any cause-of-death label. Mixed outcome extensions are desirable \citep[e.g.,][]{zhang2021large}. Second, extensions that incorporate priors over $K$ that may differ by cause can lead to posterior inference of cause-specific values of $K$. To this end, for each cause, sparsity priors over a probability simplex may be introduced to encourage absence of a subset of classes in certain domains. Third, latent class model for $\pr(\bX_i\mid Y_i,D_i)$ is a simple example of probabilistic tensor decomposition for multivariate discrete data, which can be replaced with alternatives against which comparisons are warranted \citep[e.g.,][]{bhattacharya2012simplex, zhou2015bayesian, gu2021dimension}. Fourth, negative transfer issues have been noted in machine learning literature on transfer learning \citep[e.g.,][]{pan2009survey, zhang2020overcoming}. Our approach is based on an assumption of a shared set of class-specific response profiles $\bTheta^{(c)}$ for each cause to facilitate interpretation. It is of interest to evaluate potential impact of deviations from such an assumption and to study mitigating solutions \citep[e.g.,][]{stephenson2020}. We leave these topics for future research.

% We did not adopt this approach because the nuisance nature of $K$, for which we specify a number deemed large enough during estimation. The stick-breaking specification for the class mixing weights provides some regularization towards smaller numbers of classes. 

% Investigations of negative transfer issues rely on the availability of gold-standard labels for a subset of individuals in the target domain.

% Finally, if the target domain has a substantially different conditional distribution that cannot be adequately characterized by the classes learned in the training domain, the method's performance will deteriorate. In this case, we recommend two-stage estimation, first using only training sites to estimate the class response probability profiles for each cause, and then pass these estimates to a second-stage estimation that uses all data including ones from the target domain. It is of interest to investigate how ``cut" type of operation in Bayesian inference may robustify estimations \citep[e.g.,][]{plummer2015cuts, jacob2017better,yu2021variational}.  We leaves these topics for future research.

\section*{Supplementary Materials}
Details about the main algorithm, additional simulation results, and directed acyclic graph of the model structure are presented in the Supplementary Materials.

\section*{Acknowledgments}
This work is partly supported by a seed grant from Michigan Institute of Data Science (MIDAS; to ZW, IC, ML). \textit{Conflict of Interest}: None.

%%%%%%%%%%%%%%%%%%%%%%%%%%%%%%%%%%%%%%%%%%%%%
%%%%%%% bibliography
%%%%%%%%%%%%%%%%%%%%%%%%%%%%%%%%%%%%%%%%%%%%%
\bibliographystyle{apalike}
\bibliography{double_tree}

\begin{thebibliography}{}

\bibitem[Bhattacharya and Dunson, 2012]{bhattacharya2012simplex}
Bhattacharya, A. and Dunson, D.~B. (2012).
\newblock Simplex factor models for multivariate unordered categorical data.
\newblock {\em Journal of the American Statistical Association},
  107(497):362--377.

\bibitem[Bishop, 2006]{bishop2006pattern}
Bishop, C.~M. (2006).
\newblock {\em Pattern Recognition and Machine Learning}.
\newblock Springer.

\bibitem[Blei et~al., 2017]{blei2017variational}
Blei, D.~M., Kucukelbir, A., and McAuliffe, J.~D. (2017).
\newblock Variational inference: A review for statisticians.
\newblock {\em Journal of the American Statistical Association},
  112(518):859--877.

\bibitem[Carbonetto et~al., 2012]{carbonetto2012scalable}
Carbonetto, P., Stephens, M., et~al. (2012).
\newblock Scalable variational inference for {B}ayesian variable selection in
  regression, and its accuracy in genetic association studies.
\newblock {\em Bayesian Analysis}, 7(1):73--108.

\bibitem[Chandramohan et~al., 2021]{chandramohan2021estimating}
Chandramohan, D., Fottrell, E., Leitao, J., Nichols, E., Clark, S.~J., Alsokhn,
  C., Munoz, D.~C., AbouZahr, C., Pasquale, A.~D., Mswia, R., Choi, E., Baiden,
  F., Thomas, J., Lyatuu, I., Li, Z.~R., Larbi-Debrah, P., Chu, Y., Cheburet,
  S., Sankoh, O., Bad, A.~M., Fat, D.~M., Setel, P., Jakob, R., and de~Savigny,
  D. (2021).
\newblock Estimating causes of death where there is no medical certification:
  Evolution and state of the art of verbal autopsy.
\newblock {\em In Press, Global Health Action}.

\bibitem[Datta et~al., 2021]{datta2020regularized}
Datta, A., Fiksel, J., Amouzou, A., and Zeger, S.~L. (2021).
\newblock Regularized {B}ayesian transfer learning for population-level
  etiological distributions.
\newblock {\em Biostatistics}, 22(4):836--857.

\bibitem[Dunson and Xing, 2009]{dunson2009nonparametric}
Dunson, D.~B. and Xing, C. (2009).
\newblock Nonparametric {B}ayes modeling of multivariate categorical data.
\newblock {\em Journal of the American Statistical Association},
  104(487):1042--1051.

\bibitem[Durante et~al., 2019]{durante2019conditionally}
Durante, D., Rigon, T., et~al. (2019).
\newblock Conditionally conjugate mean-field variational {B}ayes for logistic
  models.
\newblock {\em Statistical Science}, 34(3):472--485.

\bibitem[Felsenstein, 1985]{Felsenstein1985}
Felsenstein, J. (1985).
\newblock Phylogenies and the comparative method.
\newblock {\em The American Naturalist}, 125(1):1--15.

\bibitem[Fiksel et~al., 2021]{fiksel2021generalized}
Fiksel, J., Datta, A., Amouzou, A., and Zeger, S. (2021).
\newblock Generalized {B}ayes quantification learning under dataset shift.
\newblock {\em Journal of the American Statistical Association}, pages 1--19.

\bibitem[Gonz{\'a}lez et~al., 2017]{gonzalez2017review}
Gonz{\'a}lez, P., Casta{\~n}o, A., Chawla, N.~V., and Coz, J. J.~D. (2017).
\newblock A review on quantification learning.
\newblock {\em ACM Computing Surveys (CSUR)}, 50(5):1--40.

\bibitem[Grimmer, 2011]{grimmer2011introduction}
Grimmer, J. (2011).
\newblock An introduction to {B}ayesian inference via variational
  approximations.
\newblock {\em Political Analysis}, 19(1):32--47.

\bibitem[Gu et~al., 2021]{gu2021dimension}
Gu, Y., Erosheva, E.~A., Xu, G., and Dunson, D.~B. (2021).
\newblock Dimension-grouped mixed membership models for multivariate
  categorical data.
\newblock {\em arXiv preprint arXiv:2109.11705}.

\bibitem[Jaakkola and Jordan, 2000]{jaakkola2000bayesian}
Jaakkola, T.~S. and Jordan, M.~I. (2000).
\newblock Bayesian parameter estimation via variational methods.
\newblock {\em Statistics and Computing}, 10(1):25--37.

\bibitem[King and Lu, 2008]{king2008verbal}
King, G. and Lu, Y. (2008).
\newblock Verbal autopsy methods with multiple causes of death.
\newblock {\em Statistical science}, 23(1):78--91.

\bibitem[Knowles and Ghahramani, 2014]{knowles2014pitman}
Knowles, D.~A. and Ghahramani, Z. (2014).
\newblock Pitman yor diffusion trees for bayesian hierarchical clustering.
\newblock {\em IEEE transactions on pattern analysis and machine intelligence},
  37(2):271--289.

\bibitem[Koller and Friedman, 2009]{Koller2009}
Koller, D. and Friedman, N. (2009).
\newblock {\em Probabilistic graphical models: principles and techniques}.
\newblock The MIT Press.

\bibitem[Kunihama et~al., 2020]{kunihama2020bayesian}
Kunihama, T., Li, Z.~R., Clark, S.~J., McCormick, T.~H., et~al. (2020).
\newblock Bayesian factor models for probabilistic cause of death assessment
  with verbal autopsies.
\newblock {\em Annals of Applied Statistics}, 14(1):241--256.

\bibitem[Lazarsfeld, 1950]{lazarsfeld1950thelogical}
Lazarsfeld, P.~F. (1950).
\newblock {\em The logical and mathematical foundations of latent structure
  analysis}, volume~IV, chapter The American Soldier: Studies in Social
  Psychology in World War II, pages 362--412.
\newblock Princeton, NJ: Princeton University Press.

\bibitem[Li et~al., 2021a]{Li2021_treelcm}
Li, M., Park, D.~E., Aziz, M., Liu, C.~M., Price, L.~B., and Wu, Z. (2021a).
\newblock Integrating sample similarities into latent class analysis: A
  tree-structured shrinkage approach.
\newblock {\em Biometrics}, page In press.

\bibitem[Li et~al., 2020]{li2020using}
Li, Z.~R., McComick, T.~H., and Clark, S.~J. (2020).
\newblock Using {B}ayesian latent gaussian graphical models to infer symptom
  associations in verbal autopsies.
\newblock {\em Bayesian Analysis}, 15(3):781.

\bibitem[Li et~al., 2021b]{li2021openva}
Li, Z.~R., Thomas, J., Choi, E., McCormick, T.~H., and Clark, S.~J. (2021b).
\newblock The open{VA} toolkit for verbal autopsies.
\newblock {\em arXiv preprint arXiv:2109.08244}.

\bibitem[Li et~al., 2021c]{li2021bayesian}
Li, Z.~R., Wu, Z., Chen, I., and Clark, S.~J. (2021c).
\newblock {B}ayesian nested latent class models for cause-of-death assignment
  using verbal autopsies across multiple domains.
\newblock {\em arXiv preprint}.

\bibitem[McCormick et~al., 2016]{mccormick2016probabilistic}
McCormick, T.~H., Li, Z.~R., Calvert, C., Crampin, A.~C., Kahn, K., and Clark,
  S.~J. (2016).
\newblock Probabilistic cause-of-death assignment using verbal autopsies.
\newblock {\em Journal of the American Statistical Association},
  111(515):1036--1049.

\bibitem[Moran et~al., 2021]{moran2021bayesian}
Moran, K.~R., Turner, E.~L., Dunson, D., and Herring, A.~H. (2021).
\newblock Bayesian hierarchical factor regression models to infer cause of
  death from verbal autopsy data.
\newblock {\em Journal of the Royal Statistical Society: Series C (Applied
  Statistics)}.

\bibitem[Murray et~al., 2011a]{murray2011population}
Murray, C.~J., Lopez, A.~D., Black, R., Ahuja, R., Ali, S.~M., Baqui, A.,
  Dandona, L., Dantzer, E., Das, V., Dhingra, U., et~al. (2011a).
\newblock Population health metrics research consortium gold standard verbal
  autopsy validation study: design, implementation, and development of analysis
  datasets.
\newblock {\em Population Health Metrics}, 9(1):27.

\bibitem[Murray et~al., 2011b]{murray2011robust}
Murray, C.~J., Lozano, R., Flaxman, A.~D., Vahdatpour, A., and Lopez, A.~D.
  (2011b).
\newblock Robust metrics for assessing the performance of different verbal
  autopsy cause assignment methods in validation studies.
\newblock {\em Popul Health Metr}, 9(1):28.

\bibitem[Ormerod and Wand, 2010]{ormerod2010explaining}
Ormerod, J.~T. and Wand, M.~P. (2010).
\newblock Explaining variational approximations.
\newblock {\em The American Statistician}, 64(2):140--153.

\bibitem[Pan and Yang, 2009]{pan2009survey}
Pan, S.~J. and Yang, Q. (2009).
\newblock A survey on transfer learning.
\newblock {\em IEEE Transactions on knowledge and data engineering},
  22(10):1345--1359.

\bibitem[Schifeling et~al., 2016]{schifeling2016incorporating}
Schifeling, T.~A., Reiter, J.~P., et~al. (2016).
\newblock Incorporating marginal prior information in latent class models.
\newblock {\em Bayesian Analysis}, 11(2):499--518.

\bibitem[Sneath et~al., 1973]{sneath1973numerical}
Sneath, P.~H., Sokal, R.~R., et~al. (1973).
\newblock {\em Numerical taxonomy. The principles and practice of numerical
  classification.}

\bibitem[Stephenson et~al., 2020]{stephenson2020}
Stephenson, B. J.~K., Herring, A.~H., and Olshan, A. (2020).
\newblock Robust clustering with subpopulation-specific deviations.
\newblock {\em Journal of the American Statistical Association},
  115(530):521--537.

\bibitem[Thomas et~al., 2020]{thomas2019estimating}
Thomas, E.~G., Trippa, L., Parmigiani, G., and Dominici, F. (2020).
\newblock Estimating the effects of fine particulate matter on 432
  cardiovascular diseases using multi-outcome regression with tree-structured
  shrinkage.
\newblock {\em Journal of the American Statistical Association},
  115(532):1689--1699.

\bibitem[Titsias and L{\'a}zaro-Gredilla, 2011]{titsias2011spike}
Titsias, M. and L{\'a}zaro-Gredilla, M. (2011).
\newblock Spike-and-slab variational inference for multi-task and multiple
  kernel learning.
\newblock {\em Advances in Neural Information Processing Systems},
  24:2339--2347.

\bibitem[T{\"u}chler, 2008]{tuchler2008bayesian}
T{\"u}chler, R. (2008).
\newblock Bayesian variable selection for logistic models using auxiliary
  mixture sampling.
\newblock {\em Journal of Computational and Graphical Statistics},
  17(1):76--94.

\bibitem[{World Health Organization}, 2021]{world2021civil}
{World Health Organization} (2021).
\newblock Who civil registration and vital statistics strategic implementation
  plan 2021-2025.

\bibitem[Zhang et~al., 2020]{zhang2020overcoming}
Zhang, W., Deng, L., Zhang, L., and Wu, D. (2020).
\newblock Overcoming negative transfer: A survey.
\newblock {\em arXiv preprint arXiv:2009.00909}.

\bibitem[Zhang et~al., 2021]{zhang2021large}
Zhang, Z., Nishimura, A., Bastide, P., Ji, X., Payne, R.~P., Goulder, P.,
  Lemey, P., and Suchard, M.~A. (2021).
\newblock Large-scale inference of correlation among mixed-type biological
  traits with phylogenetic multivariate probit models.
\newblock {\em The Annals of Applied Statistics}, 15(1):230--251.

\bibitem[Zhou et~al., 2015]{zhou2015bayesian}
Zhou, J., Bhattacharya, A., Herring, A.~H., and Dunson, D.~B. (2015).
\newblock Bayesian factorizations of big sparse tensors.
\newblock {\em Journal of the American Statistical Association},
  110(512):1562--1576.

\end{thebibliography}

\newpage
\appendix
\begin{center}
{\Large \bf Supplementary Materials for}
\end{center}
{\Large \bf ``Tree-Informed Bayesian Multi-Source Domain Adaptation: Cross-population Probabilistic Cause-of-death Assignment using Verbal Autopsy" by Wu et al. (2021)}

%%%%%%%%%%%%%%%%%%%%%
%%%%%%%%%%%%%%%%%%%%%
%%%%%%%%%%%%%%%%%%%%%

\setcounter{equation}{0}
\renewcommand{\theequation}{A\arabic{equation}}
\setcounter{figure}{0}
\renewcommand{\figurename}{}
\renewcommand{\tablename}{}
\renewcommand{\thefigure}{Appendix Figure \arabic{figure}}
\renewcommand{\thetable}{Appendix Table \arabic{table}}
\renewcommand{\thesection}{Appendix \Alph{section}}
\renewcommand{\thesubsection}{\thesection.\arabic{subsection}}

\section{Details of the Variational Inference Algorithm}
\label{secsupp::vi}

In the following, let $q_t(A)$ represent a generic variational distribution for unknown quantities in $A$ at iteration $t$; Let $q_t(-A)$ represent the variational distribution for all but the random quantities in $A$. Let $\pr(A)$ represent a generic true joint distribution of the quantities in $A$. $[Q]:=\{1,\ldots, Q\}$ represents the set of positive integers smaller than or equal to a positive integer $Q$. The algorithm presented below deals with missing data (under missing-at-random assumption for elements of $\bX_i$ given the causes). Let $\cJ_i \subseteq \{1,\ldots, J\}$ denote the index set for the subset of observed responses for subject $i$. Let $\cI_j\subseteq \{1,\ldots,J\}$ be the index set for the subset of subjects with observed $j$-th response. Finally, recall transformed response is $X^*_{ij} = 2X_{ij}-1$; $\sigma(\bullet)$ denotes sigmoid function: $\sigma(x)=1/(1+\exp(-x))$.

\begin{enumerate}
\item[\textit{Step 0}.] Initialize the variational distribution $q_t(\cdot)$ at $t=0$. The update of each component of the variational distribution in Equation (\ref{eq:VIfamily}) of the Main Paper has a closed form that is determined by relevant first and second moments. We initialize these moments to initialize $q_0(\cdot)$. In addition, because the sigmoid functions are bounded by Gaussian kernels that depend on additional tuning parameters $(\bpsi, \bphi)$, we need to initialize them too. Finally, we initialize hyperparameters $(\btau,\btau^*)$.

In particular,
\begin{itemize}
\item Additive components of the logistic stick-breaking parameters  $\alpha_{k}^{(c,u)}$ given $s_{cu}=1$:\\ $\{(\mu_{\alpha_{k}^{(c,u)},1}, \sigma^{2}_{\alpha_{k}^{(c,u)},1}):=(E_{q_t}\{\alpha_{k}^{(c,u)}\mid s_{cu} = 1\}, V_{q_t}\{\alpha_{k}^{(c,u)}\mid s_{cu}=1\}): k\in[K-1], c\in [C], u\in \cV \}$. The mean and variance fully determine the optimal variational distribution  for $\alpha_{k}^{(c,u)}$ given $s_{cu}=1$, which can be shown to be a Gaussian distribution;
\item Logit-transformed response probabilities: $\{(\mu_{\gamma^{(u)}_{jk},1},\sigma^2_{\gamma^{(u)}_{jk},1}):=(E_{q_t}\{\gamma_{jk}^{(u)}\},V_{q_t}\{\gamma_{jk}^{(u)}\}): j\in[J], k\in[K], u\in [C]\}$; 
% \item Probability of spike-and-slab indicators: $\{p_{cu}^{(t)}=E_{q_t}[s_{cu}]: c \in[C], u\in \cV\}$ and
% $\{p_{u}^{*(t)}=E_{q_t}[s^*_{u}]: u\in \cV^*\}$; 
\item Tuning parameters in the Jaakkola-Jordan lower bounding technique: $\{\psi^{(c)}_{jk}$, $j\in[J], k\in[K]\}$, $\{\phi^{(c,g)}_k$, $c\in [C]$, $g\in\{0\}\cup[G]$, $k\in[K-1]\}$, and 
\item The hyperparameters $\{\tau_{\ell},\ell\in[L]\}$, $\{\tau^{*}_{\ell}, \ell\in[L^*]\}$.
\end{itemize}

%We emphasize that we did not restrict $q_t(\alpha_{uk})$ to the Gaussian family in Equation (14) of the Main Paper

Compute additional first and second moments as follows:
\begin{align}
%:
E_{q_t}\squared{\eta_{k}^{(c,g)}} & =\sum_{u \in a(g)}\left\{p_{cu}\left(\sigma^{2}_{\alpha_{k,1}^{(c,u)}}+(1-p_{cu})\squared{\mu_{\alpha_{k,1}^{(c,u)}}}\right)\right\} + E^2_{q_t}[\eta_{k}^{(c,g)}],\nonumber\\
E_{q_t}\squared{\alpha_{k}^{(c,u)}} &= p_{cu}\left(\sigma^{2}_{{\alpha_{k,1}^{(c,u)}}}+\squared{\mu_{\alpha_{k,1}^{(c,u)}}}\right)+(1-p_{cu})\sigma^{2}_{\alpha_{k,0}^{(c,u)}},\nonumber 
\end{align}
where $\sigma^{2}_{\alpha_{k,0}^{(c,u)}}=\tau_{\ell_u}w_u$ is the variance of $\alpha_{k}^{(c,u)}$ in its variational distribution given $s_{cu}=0$ (as will be readily seen in Step 1d below according to the VI update for $\alpha_{k}^{(c,u)}$). Similarly, for the quantities in the cause hierarchy, we compute
\begin{align}
E_{q_t}\squared{\beta_{jk}^{(c)}} &= \sum_{u \in a(c)} \sigma^{2}_{\gamma_{jk,1}^{(u)}} +E^2_{q_t}\{\beta_{jk}^{(c)}\},\nonumber \\
E_{q_t}\squared{\gamma^{(u)}_{jk}} & =   \sigma^2_{\gamma_{jk,1}^{(u)}}+\squared{\mu_{\gamma_{jk,1}^{(u)}}}.\nonumber
\end{align}

Finally, compute 
$E_{q_t}\{\eta_{k}^{(c,g)}\} = \sum_{u \in a(g)}E_{q_t}\{\xi_{k}^{(c,u)}\}$, $E_{q_t}\{\xi_{k}^{(c,u)}\}=E_{q_t}\{s_{cu} \alpha_{k}^{(c,u)}\} = \mu_{\alpha_{k,1}^{(c,u)}}$. 

$E_{q_t}[\beta_{jk}^{(c)}] = \sum_{u \in a(c)}E_{q_t}[\gamma_{jk}^{(u)}]=\sum_{u \in a(c)}\mu_{\gamma_{jk,1}^{(u)}}$.

Set initial $\cE^\ast(q)=0$.

At Step $t+1$, iterate between Step 1 to 4 until convergence (we omit iteration step index ``$t$" and ``$t+1$" in the notations below for simplicity):

\item[\textit{Step 1a}.] Update $q_{t+1}(Y_i)$ for $\{i: D_i=0\}$, by a categorical distribution with probabilities $\be_i=(e_{i1}, \ldots, e_{iC} )^\transp$:
\begin{align}
& e_{ic} \propto \exp\bigggl(E_{q_t}[\log \pi_c^{(0)}] + \sum_{k=1}^K r_{ik} F_{ik}^{(c,0)}(q_t)\bigggr), \nonumber
\end{align}
where 
\begin{align}
& F_{ik}^{(c,g)}(q_t) = \sum_{m<k}\bigggl(\log \sigma(\phi_m^{(c,g)})+\left\{-E_{q_t}(\eta_m^{(c,g)})-\phi_m^{(c,g)}\right\}/2-g(\phi_m^{(c,g)})\left\{E_{q_t}\squared{\eta_m^{(c,g)}}-\squared{\phi_m^{(c,g)}}\right\}\bigggr)\nonumber \\ 
  & + \mbf{1}\{k<K\}\bigggl(\log \sigma(\phi_k^{(c,g)})+\left\{E_{q_t}(\eta_k^{(c,g)})-\phi_k^{(c,g)}\right\}/2 - g(\phi_k^{(c,g)}) \left\{E_{q_t}\squared{\eta_k^{(c,g)}}-\squared{\phi_k^{(c,g)}}\right\} \bigggr)\nonumber \\
  & + \sum_{{\color{black} j\in \cJ_i}}\log\sigma(\psi_{jk}^{(c)})+(X^*_{ij}E_{q_t}(\beta_{jk}^{(c)})-\psi_{jk}^{(c)})/2-g(\psi_{jk}^{(c)})\left\{E_{q_t}\squared{\beta_{jk}^{(c)}}-\squared{\psi_{jk}^{(c)}}\right\},
\end{align}
for $c\in[C]$ and $g\in\{0\}\cup[G]$. In addition, for observations with observed $Y_i=c$ we set $e_{ic} = 1$ and $e_{ic'}=0$ for $c'\neq c$.

\item[\textit{Step 1b}.] Update $q_{t+1}(Z_i)$ by a categorical distribution with probabilities $\br_i = (r_{i1}, \ldots, r_{iK})^\transp$:
\begin{align}
& r_{ik} \propto \exp\bigggl(\sum_{c=1}^C e_{ic} \bigggl\{F_{ik}^{(c,g)}(q_t)\bigggr\}\bigggr).\nonumber
\end{align}
% where 
% \begin{align}
% & F_{ik}^{(\cdot,0)}(q_t) =  \sum_{m<k}\bigggl(\log \sigma(\phi_m^{(c,0)})+\left[-E_{q_t}(\eta_m^{(c,0)})-\phi_m^{(c,0)}\right]/2-g(\phi_m^{(c,0)})\left\{E_{q_t}\squared{\eta_m^{(c,0)}}-\squared{\phi_m^{(c,0)}}\right\}\bigggr)\nonumber \\ 
%   & + \mbf{1}\{k<K\} \bigggl(\log \sigma(\phi_k^{(c,0)})+\left[E_{q_t}(\eta_k^{(c,0)})-\phi_k^{(c,0)}\right]/2 - g(\phi_k^{(c,0)}) \left\{E_{q_t}\squared{\eta_k^{(c,0)}}-\squared{\phi_k^{(c,0)}}\right\} \bigggr) \nonumber \\
%   & + \sum_{j=1}^J\log\sigma(\psi_{jk}^{(c)})+\left[X_{ij}^{(\cdot,0)}E_{q_t}(\beta_{jk}^{(c)})-\psi_{jk}^{(c)}\right]/2-g(\psi_{jk}^{(c)})\left\{E_{q_t}\squared{X_{ij}^{(\cdot,0)}\beta_{jk}^{(c)}}-\squared{\psi_{jk}^{(c)}}\right\}.
% \end{align}

\item[\textit{Step 1c}.] Update $q_{t+1}(\bpi^{(g)}), g\in  \{0\}\cup[G]$ by 
\begin{align}
        q_{t+1}(\bpi^{(g)}) \propto {\sf Dirichlet}\left(\sum_{i=1}^{N}e_{i1}+d_1^{(g)}, \ldots, \sum_{i=1}^{N}e_{iC}+d_C^{(g)}\right).
\end{align}

\item [\textit{Step 1d}.] Update $q_{t+1}(s_{cu},\balpha^{(c,u)})$ for each node $u\in \cV$ of the tree $\cT$ over the $G+1$ domains, which takes a form of two-component Gaussian mixture, separately for each cause $c\in[C]$. In particular, 
\begin{align}
    & \log q_{t+1}(s_{cu},\balpha^{(c,u)}) = \EE_{q_t(-(s_{cu},\balpha\bracsup{c,u}))}\log H + \textrm{const} \nonumber \\
    & =  s_{cu} \sum_{k=1}^{K-1}\log \cN(\alpha_k^{(c,u)}; \mu_{\alpha_k^{(c,u)},1},\sigma^2_{\alpha_k^{(c,u)},1})+
    (1-s_{cu}) \sum_{k=1}^{K-1}\log \cN(\alpha_k^{(c,u)}; 0,\tau_{\ell_u}w_u) + s_{cu} \epsilon_{cu}+\textrm{const}, \nonumber 
\end{align}
where $\mu_{\alpha_k^{(c,u)},1} = D^{(c,u)}_k/C^{(c,u)}_k$, $\sigma^2_{\alpha_k^{(c,u)},1}=1/C^{(c,u)}_k$, $k\in[K-1]$. In particular,
\begin{align}
    & C^{(c,u)}_k  =\frac{1}{\tau_{\ell_u}w_u} +
    2\sum_{g \in d(u) \cap [G]} \sum_{i:Y_i=c,D_i=g} \sum_{m=k}^K r_{im} g(\phi_{k}\bracsup{c,g}) + \mathbf{1}\{0\in d(u)\}\biggl[2\sum_{i:D_i=0} e_{ic} \sum_{m=k}^K r_{im} g(\phi_{k}\bracsup{c,0})\biggr],\\
     & D^{(c,u)}_k   = \sum_{g \in d(u) \cap [G]}\sum_{i:Y_i=c,D_i=g} \bigggl[\frac{1}{2}r_{ik}-\sum_{m=k+1}^K\frac{1}{2}r_{im} - 2\left(\sum_{m=k}^K r_{im}g(\phi_{k}\bracsup{c,g})\sum_{w\in  a(g)\setminus \{u\}}E_{q_t}\{s_{cw} \alpha\bracsup{c,w}_k\}\right)\bigggr]\nonumber \\
    & + \mathbf{1}\{0\in d(u)\}\sum_{i:D_i=0} e_{ic}\bigggl[\frac{1}{2} r_{ik}-\sum_{m=k+1}^K\frac{1}{2}r_{im} - 2\left( \sum_{m=k}^K r_{im}g(\phi_{k}\bracsup{c,0})\sum_{w\in a(0)\setminus \{u\}}E_{q_t}\{s_{cw} \alpha\bracsup{c,w}_k\}\right)\bigggr]\\
   & \epsilon_{cu}  = E_{q_t}\log\frac{\rho_{c\ell_u}}{1-\rho_{c\ell_u}} + \sum_{k=1}^{K-1}\frac{\squared{D^{(c,u)}_k}}{2C^{(c,u)}_k} - \frac{1}{2}\sum_{k=1}^{K-1}\left[\log(\tau_{\ell_u}{ w_u}) + \log(C^{(c,u)}_k)\right].\label{appeq:vi_prob1}
\end{align}
It is readily seen $q(s_{cu}, \balpha^{(c,u)})$ is jointly a two-component Gaussian mixture with distinct means and variances. In particular, $q(s_{cu})$ is Bernoulli with success probability $p_{cu}=\sigma(\epsilon_{cu})$; conditional on $s_{cu}$, $q(\balpha^{(c,u)} \mid s_{cu})$ is independent Gaussians with means and variances determined by $s_{cu}$ being 1 or 0.

\item [\textit{Step 1e}.] Update $q_{t+1}(\bgamma^{(u)})$ for each node $u\in \cV^*$ of the tree $\cT^*$ over $C$ causes by 
\begin{align}
   &\log q_{t+1}(\bgamma^{(u)})  = \EE_{q_t(-\bgamma^{(u)})}\log H + \textrm{const} =\sum_{j,k} \log \cN(\gamma_{jk}^{(u)}; \mu_{\gamma_{jk}^{(u)},1},\sigma^2_{\gamma_{jk}^{(u)},1}) +\textrm{const},
\end{align}
where $\mu_{\gamma_{jk}^{(u)},1} = B_{jk}^{(u)}/A_{jk}^{(u)}$ and $\sigma^2_{\gamma_{jk}^{(u)},1} = 1/A_{jk}^{(u)}$, $j\in[J]$, $k \in[K]$. In particular, 
\begin{align}
    A_{jk}^{(u)} & = \frac{1}{\tau^*_{\ell^*_u} w^*_u}+2\sum_{c\in d(u)\cap \cC}g(\psi_{jk}^{(c)})\left(\sum_{g=1}^G  \sum_{i:Y_i=c,D_i=g}r_{ik} +\sum_{i:D_i=0}e_{ic} r_{ik}\right),\\
    B_{jk}^{(u)} & = \sum_{c\in d(u)\cap \cC} \sum_{g=1}^{G} \sum_{\color{black}i\in\{Y_i=c,D_i=g\}\cap \cI_j}\biggl\{r_{ik}X^*_{ij}/2-2r_{ik}g(\psi_{jk}^{(c)})\sum_{w\in a(c)\setminus \{u\}}E_{q_t}\{s^*_w\gamma_{jk}^{(w)}\}\biggr\}\\
    & + \sum_{c \in d(u)\cap \cC} \sum_{\color{black}i:\{D_i=0\}\cap \cI_j} e_{ic}\biggl\{r_{ik}X^*_{ij}/2-2r_{ik}g(\psi_{jk}^{(c)})\sum_{w\in a(c)\setminus\{u\}}E_{q_t}\{s^*_w \gamma_{jk}^{(w)}\} \biggr\}
    % \epsilon^*_u & = E_{q_t}\log \frac{\rho^*_{\ell^*_u}}{1-\rho^*_{\ell^*_u}} + \sum_{j=1}^J\sum_{k=1}^K \frac{\squared{B_{jk}^{(u)}}}{2A_{jk}^{(u)}} - \frac{1}{2}\sum_{j=1}^J\sum_{k=1}^K \left[\log (\tau^*_{\ell^*_u} w^*_u) + \log  A_{jk}^{(u)} \right].
\end{align}

Again it is readily seen that $q_t(\bgamma^{(u)})$ is independent Gaussians.

% \zw{Add remarks about where the information come from to inform the shrinkage: magnitude of the estimated parameters, and also relative magnitude of variance of the estimate and the $\tau_{\ell_u}$ hyperparameter of a node's parent.}

% \zw{comment on the information to estimate A, B, C, D in the Appendix; A and B uses info from all domains - and it can be viewed as LCM with an overall CSMF across domains - so should be estimated reasonably well. For C and D, the estimates are based on the information in the domain leaf only for a particular cause, so this may encounter issues regarding estimation; this is due to our choice to have $s_{cu}$ instead of $s_u$ (this needs to change the updating formula for $\epsilon_{u}$).}

\item[\textit{Step 1f}.] Update 
$$q_{t+1}(\rho_{c\ell}) = {\sf Beta}(a'_{c\ell},b'_{c\ell}), c\in[C], \ell \in [L], $$
where $a'_{c\ell} = \sum_{u\in \cV: \ell_u = \ell } E_{q_{t}}(s_{cu})+a_{c\ell}$ and $b'_{c\ell} = \sum_{u\in \cV: \ell_u = \ell }\{1-E_{q_{t}}(s_{cu})\}+b_{c\ell}$;

% Update 
% $$q_{t+1}(\rho^*_\ell) = {\sf Beta}(a''_{\ell},b''_{\ell}), \ell \in [L^*],$$
% where $a''_{\ell}=\sum_{u \in \cV^*: \ell^*_u = \ell } E_{q_{t}}(s^*_u)+a^*_\ell$ and $b''_{\ell}=\sum_{u\in \cV^*: \ell^*_u = \ell }\{1-E_{q_{t}}(s^*_u)\}+b^*_\ell$.

For every $d$ steps above, do Step 2-4:
\item[\textit{Step 2}.] Update local variational parameters $\bpsi$ and $\bphi$.
\begin{align}
    \phi\bracsup{c,g}_k = \sqrt{E_{q_{t}}\squared{\eta_{k}\bracsup{c,g}}}, \psi_{jk}\bracsup{c} = \sqrt{E_{q_t}\squared{\beta_{jk}\bracsup{c}}},
\end{align}
for $c\in[C]$, $g\in\{0\}\sqcup[G]$.

\item[\textit{Step 3}.] Update the hyperparameters $\btau$ and $\btau^*$.
\begin{align}
    \tau_\ell & = \frac{1}{C(K-1)\sum_{u\in \cV: \ell_{u}=\ell}1}\sum_{u\in \cV: \ell_{u}=\ell}\sum_{c=1}^C\sum_{k=1}^{K-1}E_{q_t}\left\{\squared{\alpha_k^{(c,u)}}/w_u\right\}, \ell \in [L],\\
    \tau^*_\ell & =  \frac{1}{JK\sum_{u\in \cV^*:\ell^*_u=\ell} 1}\sum_{u\in \cV^*:\ell^*_u=\ell}\sum_{j=1}^J\sum_{k=1}^K E_{q_t}\left\{\squared{\gamma_{jk}^{(u)}}/w^*_u\right\}, \ell \in [L^*].
\end{align}

\item[\textit{Step 4}.] Compute $\cE^\ast(q_{t+1})$ according to Appendix \ref{appendix:estar}. Stop the iteration once the absolute change in $\cE^\ast(q_{t+1})$ is less than a tolerance \verb"tol=1e-8".
The hyperparameter updates are often slower than the variational parameter updates to converge in terms of the $\cE^\ast(q_{t+1})$. In practice, we can separate the tolerance levels for the hyperparameter updates (\verb"hyper_tol=1e-4") and VI parameter updates (e.g., \verb"tol=1e-8"). One may update the hyperparameters every $d$ steps of the updates of the variational parameters. In practice, we can adjust $d$ to speed up the convergence. In this paper, we use $d=10$ which works well in simulations and data analysis. We also suggest multiple initializations to obtain a highest $\cE^\ast(q_{t+1})$ and optimal variational parameters.

\end{enumerate}

\newpage
\section{Calculation of $\log H$}
\label{secappend::logH}

Here we provide the logarithm of the lower bound $H$ for $\pr(\cD,\bGamma)$ in Equation (\ref{eq:lower_bd_prDGamma}) in the Main Paper.

\begin{align}
 \log H &   =  \sum_{g=0}^G\sum_{c=1}^C\sum_{i=1}^{N} \mathbf{1}\{Y_{i}=c,D_i=g\}\bigggl(\log \pi_c^{(g)}\\
  + & \sum_{k=1}^K \mathbf{1}\{Z_{i}=k\} \bigggl\{ \sum_{m<k}\bigggl(\log \sigma(\phi_m^{(c,g)})+(-\eta_m^{(c,g)}-\phi_m^{(c,g)})/2-g(\phi_m^{(c,g)})\left\{\squared{\eta_m^{(c,g)}}-\squared{\phi_m^{(c,g)}}\right\}\bigggr)\\ 
  + &  \mbf{1}\{k<K\}\bigggl(\log \sigma(\phi_k^{(c,g)})+(\eta_k^{(c,g)}-\phi_k^{(c,g)})/2 - g(\phi_k^{(c,g)}) \left\{\squared{\eta_k^{(c,g)}}-\squared{\phi_k^{(c,g)}}\right\} \bigggr)\\
  + &  \sum_{\color{black}j\in \cJ_{i}}\log\sigma(\psi_{jk}^{(c)})+(X^*_{ij}\beta_{jk}^{(c)}-\psi_{jk}^{(c)})/2-g(\psi_{jk}^{(c)})\left\{\squared{\beta_{jk}^{(c)}}-\squared{\psi_{jk}^{(c)}}\right\} \bigggr\}\bigggr)\\
%     & + \sum_{c=1}^C\sum_{i:D_i=0}  Y_{ic} \bigggl[\log \pi_c^{(0)}\\
%     + & \sum_{k=1}^K Z_{ik} \bigggl\{ \sum_{m<k}\bigggl(\log \sigma(\phi_m^{(c,0)})+(-\eta_m^{(c,0)}-\phi_m^{(c,0)})/2-g(\phi_m^{(c,0)})\left\{\squared{\eta_m^{(c,0)}}-\squared{\phi_m^{(c,0)}}\right\}\bigggr)\\ 
%  + &  \mbf{1}\{k<K\} \bigggl(\log \sigma(\phi_k^{(c,0)})+(\eta_k^{(c,0)}-\phi_k^{(c,0)})/2 - g(\phi_k^{(c,0)}) \left\{\squared{\eta_k^{(c,0)}}-\squared{\phi_k^{(c,0)}}\right\} \bigggr)\\
%   + &  \sum_{\color{black}j\in \cJ_{i}^{(\cdot,0)}}\log\sigma(\psi_{jk}^{(c)})+(X_{ij}^{(\cdot,0)}\beta_{jk}^{(c)}-\psi_{jk}^{(c)})/2-g(\psi_{jk}^{(c)})\left\{\squared{\beta_{jk}^{(c)}}-\squared{\psi_{jk}^{(c)}}\right\}\bigggl\}\bigggr]\\
  & +\sum_{c=1}^C \sum_{u\in \cV} \sum_{k=1}^{K-1} -\frac{1}{2}\log(2\pi \tau_{\ell_u}w_u)-\frac{1}{2\tau_{\ell_u}w_u}\squared{\alpha_{k}^{(c,u)}}\\
  &+\sum_{u\in \cV^*}\sum_{j=1}^J\sum_{k=1}^K -\frac{1}{2}\log(2\pi\tau^*_{\ell^*_u}w^*_u)-\frac{1}{2\tau^*_{\ell^*_u}w^*_u}\squared{\gamma_{jk}^{(u)}}\\
  & + \sum_{c=1}^C\sum_{u\in \cV}\left[ {s_{cu}} \log \rho_{c\ell_u}+ (1-s_{cu})\log(1-\rho_{c\ell_u})\right]\\%+\sum_{u\in \cV^*} %\left[{s^*_u} \log \rho^*_{\ell_u}+ (1-s^*_u)\log(1-\rho^*_{\ell_u})\right]\\
  &+\sum_{c=1}^C\sum_{\ell=1}^L\left[ (a_{c\ell}-1)\log \rho_{c\ell}+(b_{c\ell}-1)\log(1-\rho_{c\ell})-\log {\sf B}(a_{c\ell},b_{c\ell})\right]\\
%   &+\sum_{\ell=1}^{L^*}\left[ (a^*_\ell-1)\log \rho^*_\ell+(b^*_\ell-1)\log(1-\rho^*_\ell)-\log {\sf B}(a^*_{\ell},b^*_{\ell})\right]\\
  & + \sum_{g=0}^G \sum_{c=1}^C (d_c^{(g)}-1)\log \pi_c^{(g)}+\mathrm{const},
\end{align}
where $\textrm{const}$ is a term that does not depend on $\bGamma$.

\section{Calculation of $\cE^*(q)$}
\label{appendix:estar}

For ease of presentation, we omit the iterator $t$ in the following. We have $\cE^*(q)   = E_{q} \log(H) - E_{q}\log q + \textrm{const}$, where the two non-constant terms are:
\noindent
\begin{align}
  E_{q} \log(H) = &   \sum_{g=0}^G\sum_{c=1}^C\sum_{i=1}^{N}  e_{ic} \bigggl\{E_q [\log \pi_c^{(g)}]+\sum_{k=1}^K r_{ik}F_{ik}^{(c,g)}(q)\bigggr\}\\
  +& \sum_{c=1}^C \sum_{u\in \cV} \sum_{k=1}^{K-1} -\frac{1}{2}\log(2\pi \tau_{\ell_u}w_u)-\frac{1}{2\tau_{\ell_u}w_u}E_q\squared{\alpha_{k}^{(c,u)}}\\
  +&\sum_{u\in \cV^*}\sum_{j=1}^J\sum_{k=1}^K -\frac{1}{2}\log(2\pi\tau^*_{\ell^*_u}w^*_u)-\frac{1}{2\tau^*_{\ell^*_u}w^*_u}E_q\squared{\gamma_{jk}^{(u)}}\\
  +&  \sum_{c=1}^C\sum_{u\in \cV} E_q\{s_{cu}\} E_q\log \rho_{c\ell_u}+ (1-E_q \{s_{cu}\})E_q\log(1-\rho_{c\ell_u})\\
%   +&\sum_{u\in \cV^*} E_q[s^*_u] E_q\log \rho^*_{\ell_u}+ (1-E_q [s^*_u])E_q\log(1-\rho^*_{\ell_u})\\
  +&\sum_{c=1}^C\sum_{\ell=1}^L (a_{c\ell}-1)E_q \log \rho_{c\ell}+(b_{c\ell}-1)E_q\log(1-\rho_{c\ell})-{\log {\sf Beta}(a_{c\ell},b_{c\ell})}\\
%   +&\sum_{\ell=1}^{L^*} (a^*_\ell-1)E_q\log \rho^*_\ell+(b^*_\ell-1)E_q\log(1-\rho^*_\ell) - {\log {\sf Beta}(a^*_{\ell},b^*_{\ell})}\\
  + & \sum_{g=0}^G \sum_{c=1}^C (d_c^{(g)}-1)E_q\log \pi_c^{(g)} - {\sum_{g=0}^G\log {\sf B}(\bd^{(g)})},
  \end{align}
  where ${\sf B}(\bx = (x_1, \ldots, x_I)) = {\prod_i \Gamma(x_i)}/{\Gamma(\sum_i x_i)}$ and $\Gamma(\cdot)$ is the Gamma function, $x_i >0, i\in[I]$ (when $I=2$, $B(\cdot)$ is the Beta function); and 
  
  \begin{align}
 - E_{q}\log q = - &  \sum_{g=0}^G  \left(\sum_{c=1}^C \left(\sum_{i=1}^{N}e_{ic}+d^{(c,g)}-1\right)E_q\{\log (\pi^{(\cdot,g)}_c)\}-\log {\sf B}(\sum_{i=1}^{N}e_{ic}+d^{(c,g)},c=1,\ldots, C)\right)\\
%   -&  \sum_{c=1}^C\sum_{u\in \cV} E_q \log q(s_{cu},\balpha^{(c,u)})-  \sum_{u\in \cV^*} E_q \log q(s^*_u,\bgamma^{(u)})\zw{need details}\\
  +& {0.5}\sum_{c=1}^C\sum_{u\in \cV}\sum_{k=1}^{K-1}E_q\{s_{cu}\}+E_{q}\{s_{cu}\}\log(2\pi \sigma^2_{\alpha_{k}^{(c,u)},1})\\
  +& {0.5}\sum_{c=1}^C\sum_{u\in \cV}\sum_{k=1}^{K-1}E_q\{1-s_{cu}\}+E_{q}\{1-s_{cu}\}\log(2\pi \tau_{\ell_u}w_u)\\
%   +& {0.5}\sum_{u\in \cV^*}\sum_{j=1}^J\sum_{k=1}^K E_q[s^*_u]+E_q[s^*_u]\log(2\pi \sigma^2_{\gamma_{jk}^{(u)},1})\\
%   +& {0.5}\sum_{u\in \cV^*}\sum_{j=1}^J\sum_{k=1}^K E_q[1-s^*_u]+E_q[1-s^*_u]\log(2\pi \tau^*_{\ell^*_u} w^*_u)\\
  -&  \sum_{i:D_i=0} \sum_{c=1}^C e_{ic} \log e_{ic} -  \sum_{i=1}^{N} \sum_{k=1}^K r_{ik} \log r_{ik}\\
  -&\sum_{c=1}^C\sum_{u\in \cV} \left\{E_q[s_{cu}]\log(p_{cu})+E_q[1-s_{cu}]\log(1-p_{cu})\right\}\nonumber\\
%   - &  \sum_{u\in \cV^*} \left\{E_q[s^*_{u}]\log(p^*_{u})+E_q[1-s^*_{u}]\log(1-p^*_{u})\right\}\\
  - & \sum_{c=1}^C\sum_{\ell=1}^L \biggl\{(a'_{c\ell}-1)E_q \{\log \rho_{c\ell}\} + (b'_{c\ell}-1) E_{q}\{\log (1-\rho_{c\ell})\} - \log {\sf B}(a'_{c\ell},b'_{c\ell})\biggr\}
    % - &  \sum_{\ell=1}^{L^*} \biggl[(a''_\ell-1)E_q [\log \rho^*_\ell] + (b''_\ell-1) E_{q}[\log (1-\rho^*_\ell)]  - \log {\sf B}(a''_\ell,b''_\ell)\biggr].
\end{align}

\section{Additional Details of Simulation Studies}
\label{secappendix::simulation_details}

\paragraph{Simulation I} We use the domain hierarchy with $\pleaf=6$ domain leaves and $2$ non-root nodes with root node $u=u_0=1$ (see Figure \ref{fig::simI_tree_setup} in the Main paper). The total number of causes is $C=3$. We set the total sample sizes to be $N=1000$ with the domain-specific sample sizes being 1) evenly and randomly distributed across domains or 2) unevenly and randomly allocated by domain: we first form pairs of domains and evenly and randomly allocated the total sample sizes to all the pairs of domains; then within each pair, we randomly allocate samples with a ratio of 4 to 1. In addition, we set $G=5$ source domains and $1$ target domain; the number of latent classes for each cause is $K=2$, for $J=20,60$ binary responses. We considered two scenarios of the response probability profiles: 1) stronger signal: $\theta_{j1}^{(c,g)}=0.95$, $\theta_{j2}^{(c,g)}=0.05$; 2) weaker signal: $\theta_{j1}^{(c,g)}=0.8$, $\theta_{j2}^{(c,g)}=0.2$. 

Two scenarios of between-domain patterns of CSMFs are considered: 1) balanced: $\pi^{(g)}_c = 1/C$, and 2) unbalanced: $\bv^{(g)}= (x_1,x_2,\ldots, x_{C})/C$ and $x_c = 5$ if $c=1$, and $x_c=3$ if $c \not \equiv 0 \;(\bmod\; C)$, $c=1, \ldots C$. We picked the target domain CSMF to be $\bpi^{(0)} = \bv^{(3)}$ and $\bpi^{(g)}$, $g=1, \ldots, G$ to take the rest of vectors: $\bv^{(1)}, \bv^{(2)}, \bv^{(4)}, \ldots, \bv^{(G)}$. 

For each domain, the class mixing weights $\blambda^{(c,g)}$ are generated independently for each cause $c$ by the following scheme: 1) for cause $c$, sample independently $\balpha^{(c,u_0)}$ for the root domain node: $\balpha^{(c,u_0)}\sim { \bF}({\sf Dirichlet}(2,K))$, where $\bF$: $\cS^{K-1}\rightarrow \RR^{K-1}$  maps a vector in the $K$-probability simplex to a vector in the $K-1$ dimensional Euclidean space $\bF(\blambda)=\balpha$ where $\balpha=(\alpha_1,\ldots, \alpha_{K-1})$ is the unique vector that satisfies $\lambda_1 = \sigma(\alpha_1)$, \ldots, $\lambda_k = \sigma(\alpha_k)\prod_{s<k}(1-\sigma(\alpha_s))$, $\ldots$, and $\lambda_K = \prod_{s<K}(1-\sigma(\alpha_s))$; 2) For cause $c$, set the same and fixed diffusions upon $\balpha^{(c,u)}$ for non-root nodes $u$ to be $-2$ if $u=2$, $2$ if $u=3$, and zero for $u\geq 3$.

The simulation setup creates a scenario the {\sf True Domain Grouping} of four blocks: \{{\sf 0, 1}\}, and \{{\sf 2, 3}\}, \{{\sf 4}\}, \{{\sf 5}\}, . The {\sf Complete Pooling} approach sets $s_{cu}=0$ for any non-root node $u\in \cV\setminus u_0$, forming a single group of six domains. The {\sf Ad Hoc Domain Grouping} method splits \{{\sf 0, 1}\} into \{{\sf 0}\} and \{{\sf 1}\} resulting in a finer domain grouping. For the {\sf No Domain Grouping} approach, we share the class-specific response profiles, but do not borrow information across domains to perform shrinkage about the mixing weights $\blambda^{(c,g)}$, $g\in \{0\}\cup [G]$. In the method {\sf Domain Adaptive}, we used hyperparameters $a_{c\ell}=b_{c\ell}=1$ in the selection probability of the spike-and-slab prior along the domain hierarchy.  For all approaches, we set $\bd^{(g)} = (1,\ldots, 1)$ for all the domains. During estimation, we use a two-level cause tree with a root node pointing towards $C$ cause leaves with equal edge weights.

% \label{sec:semisynthetic_design}
\subsection{Simulation IIa and IIb}
\label{secapp:simulationII}
\paragraph{Design}
Two designs (referred to as IIa and IIb) are considered where the difference lies in how the masked CODs are chosen, in a uniform or non-uniform fashion over the causes. 

In Simulation IIa), we randomly split subjects into $80\%$ training and $20\%$ testing data. We then collect the $20\%$ split from each PHMRC site into a single target domain on which CSMF and CODs are to be inferred. In this basic setup, the causes-of-deaths in the target domain are close to the population average across domains; the conditional distributions of the VA responses given the cause is also close to the counterpart estimates based on data from the source domains. 

In Simulation IIb) for each cause, we draw a random fraction of deaths $\varphi_c$ generated from a half-half mixture: $0.5{\sf Beta}(1,5)+0.5{\sf Beta(1,20)}$; $\varphi_c$ is also independently generated across causes, so that when constructing the target domain data some causes are up-sampled while others are down-sampled relative to the global CSMFs. We have designed such a scheme to let the constructed target domain to have a CSMF that is different from those in the source domains. We then collect the sampled deaths into a single domain, and treat it as the target domain on which the CSMFs and CODs are to be inferred. In this setup, the target may have distinct CSMFs from other domains; the conditional distribution of VA responses given a cause is a mixture across the other domains. In both cases, the domain trees have $\pleaf=7$ leaves and $p-\pleaf=3$ non-leaf nodes. Note that because the constructed target domain is a random sample from the entire data, we specify weights for the edges in the tree so that the tree-based distance from the constructed target domain to the six original domains are identical.

% 2) methods that do not perform domain adaptation (e.g., ignores sites from the same country; the tree set it as a prior in terms of two sibling leaves in the domain tree); 3) methods that do not perform cause tree structured shrinkage - either they cannot estimate or require manual specification of groupings. 

\paragraph{Results} Figure \ref{fig:simulationII} shows the relative comparisons of the various options of conducting target domain CSMF estimation in terms of CMSF accuracy; unlike Simulation I, here the {\sf True Domain Grouping} comparator is unavailable. In particular, the domain adaptive method which adaptively encourage shrinkage along the domain hierarchy produced estimates with slightly better accuracy overall. In addition, the task of CSMF estimation in the constructed target domain is more challenging when CSMFs differ substantially from the source domains. 

\begin{figure}[H]
\captionsetup{width=0.9\linewidth}
\centering
\addtocounter{figure}{0}
{\subfigure[CSMF Accuracy Comparison]{
\includegraphics[width=.5\linewidth]{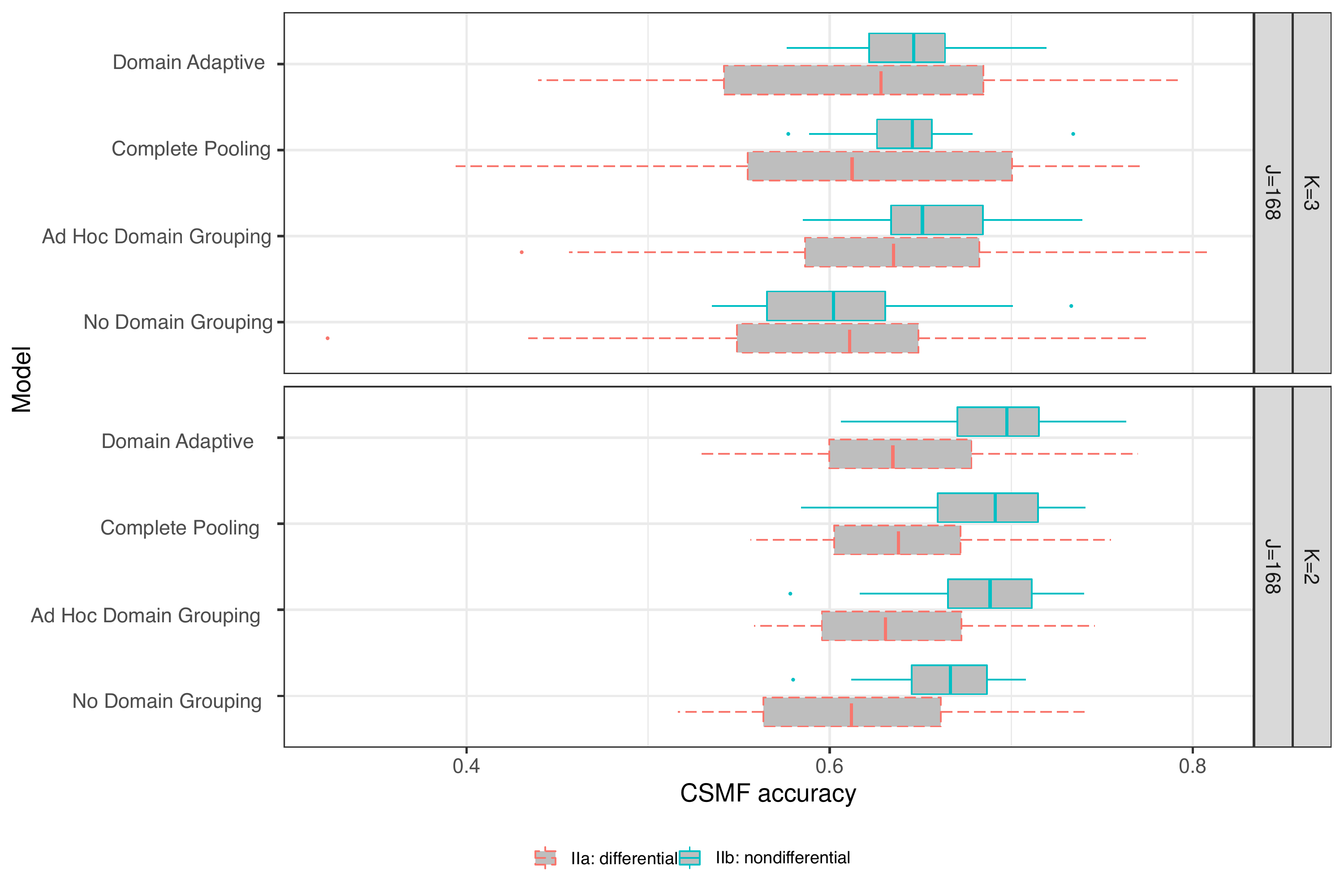}
\label{fig:simII_csmf_acc}
}
}
\addtocounter{figure}{0}
\caption{Simulation IIa and IIb show the proposed method achieves better estimation accuracy in terms of CSMF accuracy.}
\label{fig:simulationII}
\end{figure}

\section{Tree-Structured Shrinkage Priors: A Review}
\label{sec:tree_shrinkage_overview}

% \zw{Insert the overall strategy by using a generic notation; then say we specialize to $\bTheta^{(c)}$ and $\blambda$s.}

% \zw{comment on the ability to let the leaves in the same clade to have distinct parameter values.}

% \zw{tree length discussion}.

We specify a prior distribution for a set of real-valued parameters without range constraints that may differ by leaf nodes $\{\vartheta_v: v \in \vleaf\}$. In specifying the tree-structure shrinkage prior, we need a few pieces of tree-related information: a weighted rooted tree $\cT_w = (\cT=(\cV,E),w)$ with leaves $\vleaf\subset \cV$, edge lengths $\bw = (w_u)_{u\in \cV}$, the leaf id for each observation $\cL = (v_1,\ldots, v_N)^\transp$ where the sample-to-leaf indicator $v_i$ chooses parameter $\vartheta_{v_i}$ to partly characterize the distribution of data from subject $i$. Because leaf-specific sample sizes may vary, we propose a tree-structured prior to borrow information across nearby leaves. The prior encourages collapsing certain parts of the tree so that observations within a collapsed leaf group share the same parameter value.  \cite{Li2021_treelcm} has extended \cite{thomas2019estimating} to deal with rooted weighted trees.

We specify a spike-and-slab Gaussian diffusion process prior along a rooted weighted tree for $\vartheta_v$. For a leaf $v\in \vleaf$, let
\begin{align}
\vartheta_{v} & = \sum_{u\in a(v)}\varphi_{u}. \label{eq:tree_shrinkage}
\end{align}
Here $\vartheta_{v}$ is defined for leaves only and $\varphi_{u}$ is defined for all the nodes. Suppose $v$ and $v'$ are leaves and siblings in the tree such that $pa(v) = pa(v')$, setting $\varphi_{v} = \varphi_{v'} = 0$ implies $\vartheta_{v} = \vartheta_{v'}$. More generally, a sufficient condition for $M$ leaves $\vartheta_{v}$, $v\in \{v_1, \ldots, v_{M}\}$ to fuse is to set $\varphi_{u}=0$ for any $u$ that is an ancestor of any of $\{v_1, \ldots, v_M\}$ but not common ancestors for all $v_m$. That is, to achieve grouping of observations that share the same vector of latent class proportions, in our model, it is equivalent to parameter fusing. In the following, we specify a prior on the $\varphi_{u}$ that {\it a priori} encourages sparsity, so that closely related observations are likely grouped to have the same vector of class proportions. The fewer distinct ancestors two nodes have, the more likely the parameters $\vartheta_{v}$ are fused, because the prior would encourage fewer auxiliary variables $\varphi_{u}$ to be set to zero. In particular, we specify
\begin{align}
\varphi_{u} & =  s_{u} \alpha_{u}, \forall ~u\in \cV,\label{eq:ss_pi}\\
\alpha_{u}  & \sim N(0,\tau_{\ell_u}w_\pau), ~\text{independently~for}~\forall ~u\in \cV, \label{eq:alpha_norm}\\
s_{u_0}=1, \text{~and~} s_u & \sim {\sf Bernoulli}(\varrho_{\ell_u}), ~\text{independently~for}~u\in \cV\setminus u_0, \label{eq:ssindicator}\\
\varrho_\ell & \sim {\sf Beta}(a_\ell, b_\ell), \textrm{~independently~for~}\ell \in [L], \label{eq::hyper_prior_with_level0}
\end{align}
where $N(m,s)$ represents a Gaussian density function with mean $m$ and variance $s$. $\tau_{\ell}$ is the unit-length variance and controls the degree of diffusion along the tree which may differ by node level $\ell_u$ where $\ell_u\in [L]$ represents the ``level" or ``hyperparameter set indicator" for node $u$. For example, in simulations and data analysis, we will assume that the root for the diffusion process has a prior unit-length variance distinct from other non-root nodes. For the root $u_0$ with $s_{u_0}=1$, $\alpha_{u_0}$ initializes the diffusion of $\vartheta_{u}$. 

Leaf groups are formed by selecting a subset of nodes in $\cV$: $\cU=\{u\in \cV: s_u=1\}$. Except a probability-zero set, two leaves $v$ and $v'$ are grouped, or ``fused", if and only if $a(v)\cap \cU=a(v')\cap \cU$. In particular, the null set is $\{\vartheta_{v}=\vartheta_{v'}\}\cap\{\sum_{u\in [a(v)\cap \cU] \setminus [a(v')\cap \cU] }\alpha_{u}=\sum_{u\in [a(v')\cap \cU] \setminus [a(v)\cap \cU] }\alpha_{u}\}$ where the latter has probability zero. We may estimate $\cU$, e.g., using the posterior median model.

% Smaller values of $\varrho_{\ell_u}$ encourage parameter fusion of sibling nodes.  

\begin{remark}
Equations (\ref{eq:tree_shrinkage})-(\ref{eq::hyper_prior_with_level0}) define a Gaussian diffusion process initiated at $\alpha_{u_0}$:
\begin{align}
\vartheta_{u} \mid \{\varphi_{u'},u\in a(u)\}, s_u, \tau_{\ell_u}, w_{\pau} & \sim N\left(\sum_{u'\in a(u)}\xi_{u'}, s_u\tau_{\ell_u}w_{\pau}\right), 
\end{align}
for any non-root node $u\neq u_0$; also see the seminal formulation by \citet{Felsenstein1985}. To aid the understanding of this Gaussian diffusion prior, it is helpful to consider a special case of $s_u=1$ and $\ell_u = 1$, $\forall u \in \cV$. For two leaves $v,v'\in \vleaf$, the prior correlation between $\vartheta_{v}$ and $\vartheta_{v'}$ is 
\begin{align}
\Cor(\vartheta_{v}, \vartheta_{v'}) & = \frac{\sum_{u\in a(v)\cap a(v')}w_{\pau}}{\left\{dist_{\cT_w}(u_0,v)dist_{\cT_w}(u_0,v')\right\}^{1/2}},
\end{align}
When $v$ and $v'$ have the same number of ancestors $(|a(v)| = |a(v')|)$ and all edges have identical weight $w_u=c, \forall u$, the prior correlation is the fraction of common ancestors. 
\end{remark}

\section{Appendix Figures}

\begin{figure}[H]
\captionsetup{format=plain,width=0.9\linewidth}
\centering
\includegraphics[width=0.95\textwidth]{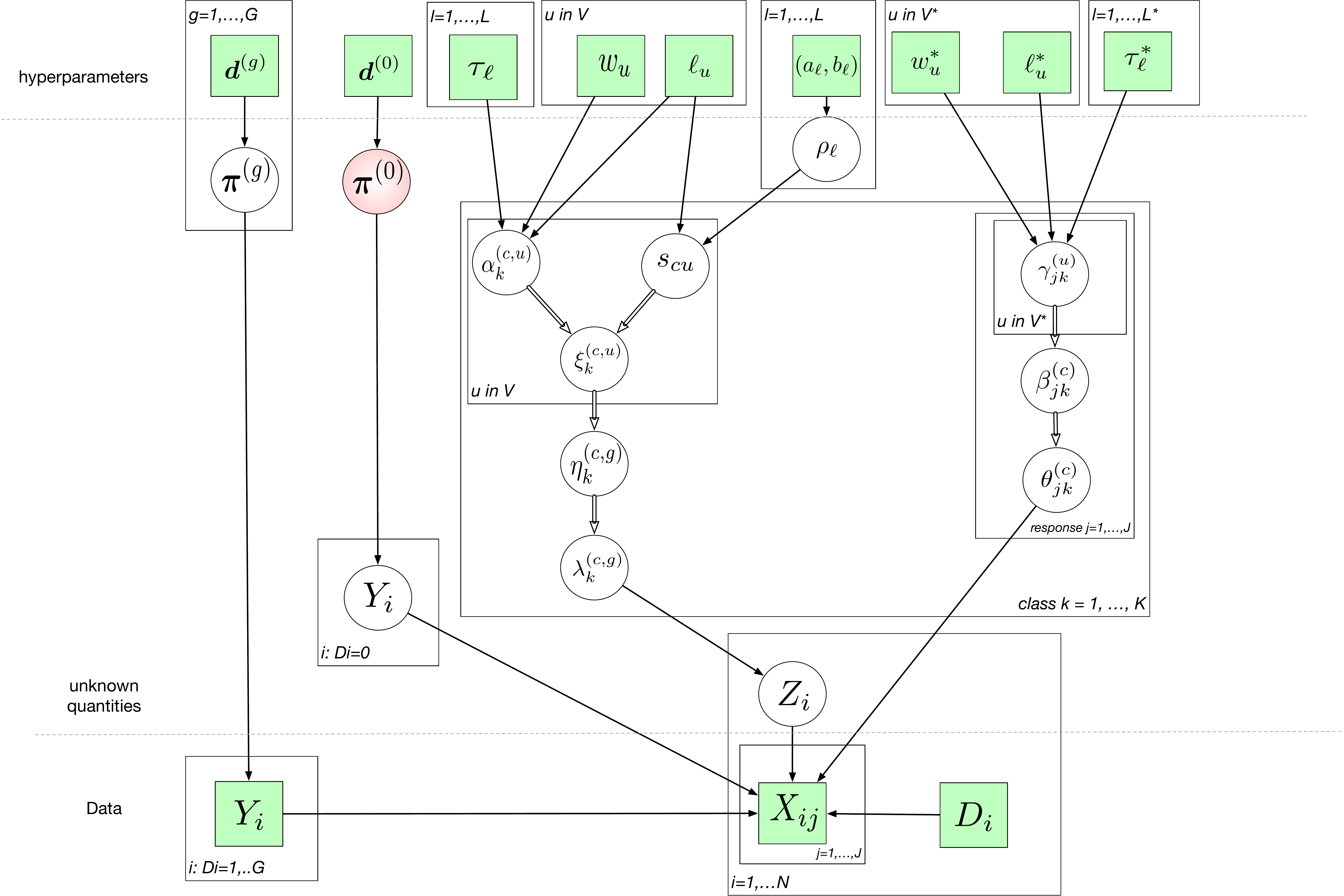}
\caption{The directed acyclic graph (DAG) representing the structure of the model likelihood and priors following the style of \citet{Koller2009}. The quantities in squares are either data or hyperparameters; the unknown quantities are shown in the circles; the double-stroke circle $Z_i$ indicates a selector, choosing the latent class $k=1,\ldots,K$. The arrows connecting variables indicate that the parent parameterizes the distribution of the child node (solid lines) or completely determines the value of the child node (double-stroke arrows). The rectangular ``plates" where the variables are enclosed indicate that a similar graphical structure is repeated over the index; The index in a plate indicates nodes, hyperparameter levels, leaves, subjects, classes and features. The parameter of interest $\bpi^{(0)}$, the CSMFs in the target domain, is highlighted.}
\label{fig:doubletree_model_dag}
\end{figure}

\begin{figure}[H]
\captionsetup{width=0.95\textwidth}
\centering
\includegraphics[width=.9\textwidth]{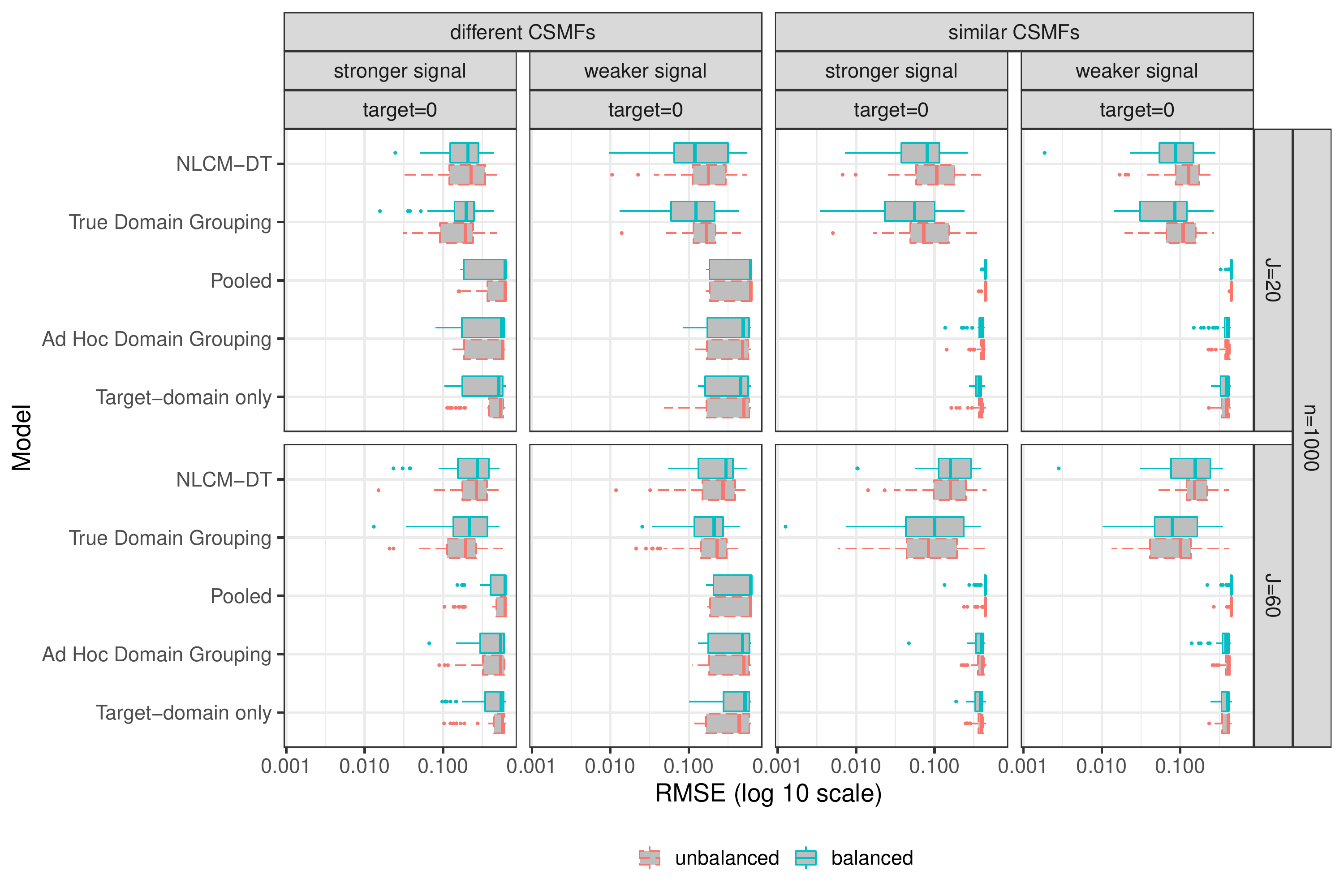}
\caption{Simulation I: RMSE comparison.} 
\label{fig::simI_RMSE}
\end{figure}

\end{document}